\documentclass{article}
\usepackage[utf8]{inputenc}

\def\includegraphics{}
\usepackage{graphicx}
\usepackage{subcaption}
\usepackage[normalem]{ulem}
\usepackage{authblk}
\usepackage{amsmath,amssymb}
\usepackage[numbers,sort]{natbib}

\title{Heterogeneous network approach to predict individuals' mental health}

\author[1]{Shikang Liu}
\author[1]{Fatemeh Vahedian}
\author[2]{David Hachen}
\author[3]{Omar Lizardo}
\author[1]{Christian Poellabauer}
\author[1]{Aaron Striegel}
\author[1,4,5,*]{Tijana Milenkovi\'{c}}

\affil[1]{Department of Computer Science and Engineering, University of Notre Dame}
\affil[2]{Department of Sociology, University of Notre Dame}
\affil[3]{Department of Sociology, University of California, Los Angeles}
\affil[4]{Eck Institute for Global Health, University of Notre Dame}
\affil[5]{Center for Network and Data Science (CNDS), University of Notre Dame}
\affil[*]{Corresponding author (email: tmilenko@nd.edu)}

\date{ }

\begin{document}

\maketitle

\begin{abstract}
Depression and anxiety are critical public health issues affecting millions of people around the world. To identify individuals who are vulnerable to depression and anxiety, predictive models have been built that typically utilize data from one source. Unlike these traditional models, in this study, we leverage a rich heterogeneous data set from the University of Notre Dame's NetHealth study that collected individuals' (student participants') social interaction data via smartphones, health-related behavioral data via wearables (Fitbit), and trait data from surveys. To integrate the different types of information, we model the NetHealth data as a heterogeneous information network (HIN). Then, we redefine the problem of predicting individuals' mental health conditions (depression or anxiety) in a novel manner, as applying to our HIN a popular paradigm of a recommender system (RS), which is typically used to predict the preference that a person would give to an item (e.g., a movie or book). In our case, the items are the individuals' different mental health states. We evaluate four state-of-the-art RS approaches. Also, we model the prediction of individuals' mental health as another problem type - that of node classification (NC) in our HIN, evaluating in the process four node features under logistic regression as a proof-of-concept classifier. We find that our RS and NC network methods produce more accurate predictions than a logistic regression model using the same NetHealth data in the traditional \emph{non-network} fashion as well as a random-approach. Also, we find that the best of the considered RS approaches outperforms all considered NC approaches. This is the first study to integrate smartphone, wearable sensor, and survey data in an HIN manner and use RS or NC on the HIN to predict individuals' mental health conditions. 

\end{abstract}

\section{Introduction}

Mental disorders such as depression and anxiety have been recognized as critical public health issues. Mental disorders are one of the leading causes of both injury and disability for people around the world \cite{wahlbeck2015public,cassano2002depression}. According to the World Health Organization (WHO), in 2015, 322 million people were living with depression (4.4\% of the global population) and 264 million people were living with anxiety (3.6\% of the global population) \cite{world2017depression}. Moreover, depression and anxiety are major contributors to suicides \cite{world2017depression}. Although early interventions significantly reduce the risk of developing mental disorders, about two-thirds of people do not seek appropriate treatments due to a lack of awareness of their mental illness \cite{berk2011does,aldarwish2017predicting}. One way to overcome this issue is to develop predictive models to enable individuals to recognize their risks of mental disorders. Fortunately, smartphones, wearable sensors, and online social media provide a wealth of data relevant to individuals' mental health \cite{glenn2014new,sano2015recognizing,de2013predicting}. Using such data, health care providers could alert people who are at risk for depression and anxiety to get timely treatment.

In terms of developing models to predict mental health, existing studies can be categorized into three different groups according to the types of data that they use \cite{mohr2017personal, glenn2014new}: 1) one group of studies rely on \textit{smartphone usage data} \cite{ bogomolov2014daily,puiatti2011smartphone, alvarez2014tell}, such as incoming and outgoing call frequency and duration; 2) another group of studies use \textit{wearable sensor data}, such as physical activity, skin conductance, and heart rate \cite{vallance2011associations,picard2016multiple,gravenhorst2013towards}; and 3) the remaining group of studies use \textit{social media behavioral data}, such as text or image content on social media platforms \cite{wongkoblap2017researching,de2013predicting,aldarwish2017predicting}.

However, existing studies have several limitations. First, among group 1 and group 2 studies, most are conducted on a limited number of individuals (fewer than 50) and a limited time period (less than one month) \cite{mohr2017personal} except for one that collects smartphone data from 111 individuals for seven months \cite{bogomolov2014daily}. Critically, the small number of individuals may not be representative of a larger population and thus the obtained results might not be generalizable. In addition, data collected from a short time period may be affected by special events such as holidays and thus the obtained results might not be reliable. Group 3 studies do not suffer from a limited number of individuals or a short time period because data is collected online, but the prediction performance of their models may be affected by low data quality resulting from fake accounts and noise in text and image contents \cite{glenn2014new,wongkoblap2017researching}. Second, all of the existing studies focused on a single type of data except for two \cite{sano2015recognizing,bogomolov2014daily}. Models using a single type of data are often less accurate than models integrating multiple types of data \cite{bogomolov2014daily,sano2015recognizing}. For example, it was shown that combining different types of data including information on individuals' personality traits, weather conditions, and smartphone data yields more accurate performance in predicting individuals' stress levels than individual data types \cite{bogomolov2014daily}. Third, among the existing studies, only three \cite{wang2013improved,lin2018social,psb2020} have utilized a network approach to model and analyze data by capturing relationships between entities. Clearly, there is a shortage of network-based methods that could utilize relations between entities to make mental health predictions \cite{wongkoblap2017researching}. Developing new approaches of this type is essential because networks are powerful models of complex real-world systems, including social networks, and because social networks play an important role in individuals' mental health conditions \cite{rosenquist2011social,smith2008social,schaefer2011misery,meng2016interplay}. Moreover, networks allow for data integration in an elegant way and can be used to make predictions about individuals' health-related traits \cite{hosseini2018heteromed,zitnik2018modeling}. For instance, a network approach was used to model multiple types of clinical data of patients in order to diagnose diseases \cite{hosseini2018heteromed}.

To address the limitations of the existing studies, we propose the following contributions.

We leverage a rich data set from the NetHealth study to predict individuals' mental health conditions. The NetHealth study collected smartphone data, wearable sensor (Fitbit) data, and individuals' trait data from surveys from approximately 700 undergraduate student participants at the University of Notre Dame during 2015 to 2019 \cite{liu2018network,purta2016experiences, faust2017exploring}. The NetHealth data is more representative of a large population since it contains a larger number of individuals (student participants) than previous studies. In addition, the NetHealth data covers a longer time period than the data from the existing studies and thus may be more reliable. Moreover, the NetHealth study contains multiple types of data for \textit{the same set of individuals}, which allows us to have a more comprehensive and thus hopefully more accurate understanding of the individuals' mental health conditions compared to the existing studies that only focused on a single type of data.

\begin{figure}[h!]
\includegraphics[width=\linewidth]{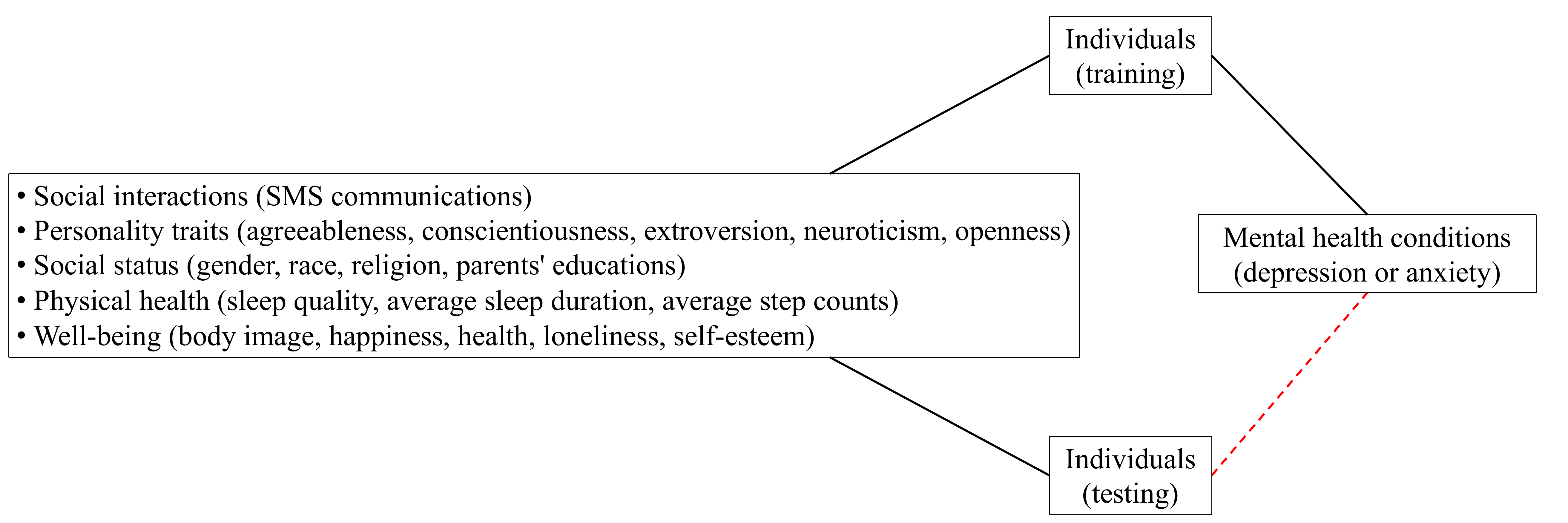}
\captionsetup{width=\textwidth}
\caption{The summary of our data and the goal of our study.}
\label{fig:3}
\end{figure}

In this study, we aim to predict individuals' mental health conditions by integrating multiple types of data including individuals' social interactions, i.e., their SMS communications collected from smartphones, health-related behaviors collected via Fitbits, and individuals' traits (e.g., personalities) collected from surveys (Figure \ref{fig:3}). We divide these data collected from the three sources (smartphones, Fitbits, and surveys) into five dimensions: individuals' social interactions (i.e., SMS communications), personality traits, social status, physical health, and well-being, with each of the last four dimensions consisting of several components (Figure \ref{fig:3}). To predict individuals' mental health conditions, we first divide individuals into training and testing sets. In the prediction task, we leverage all individuals' information as shown in the left box of Figure \ref{fig:3} as well as the training individuals' mental health conditions to train a predictive model. Then we use the trained model to predict the testing individuals' mental health conditions.

\begin{figure}[h!]
\includegraphics[width=0.65\linewidth]{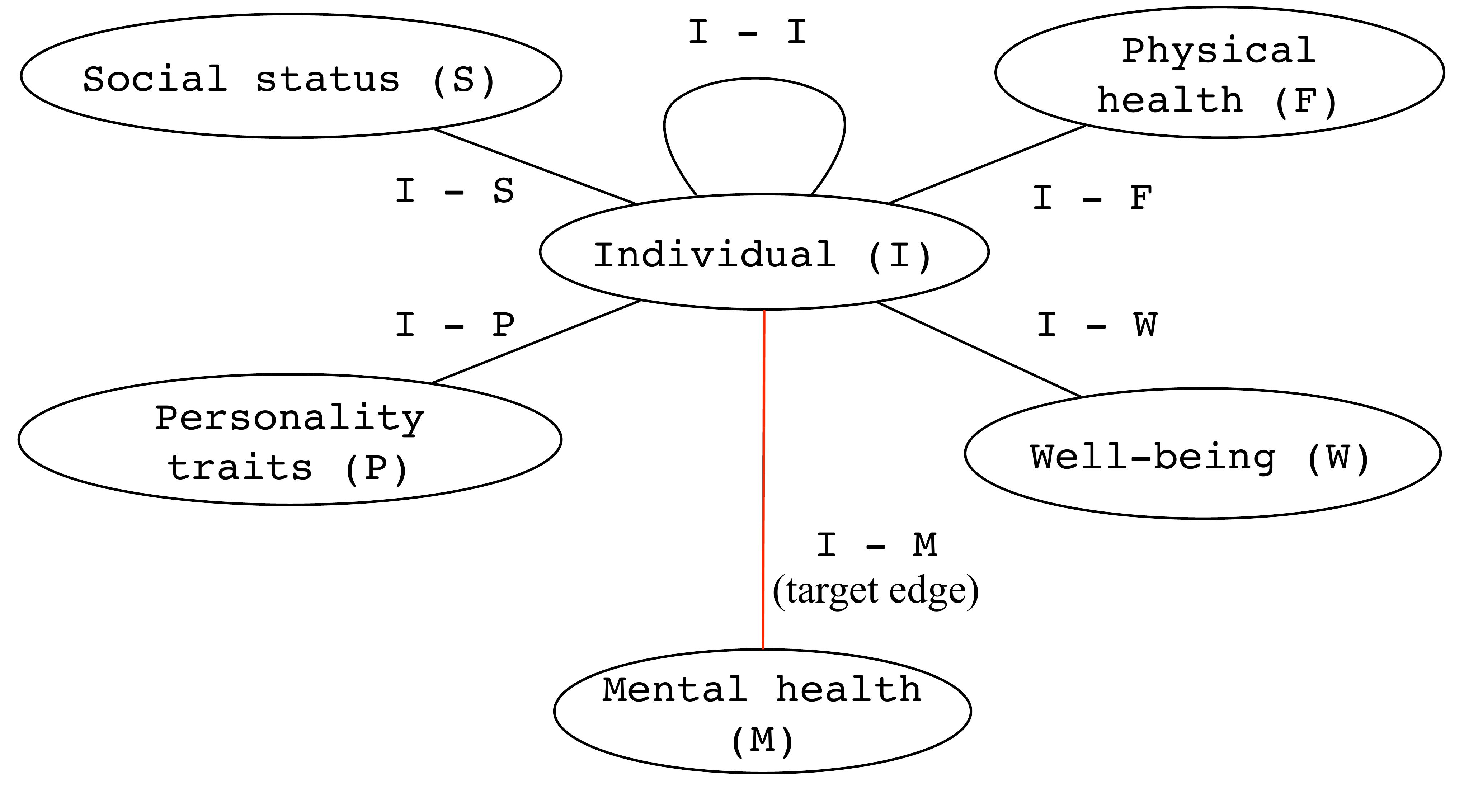}
\captionsetup{width=\textwidth}
\caption{Network schema of the HIN that we construct from the NetHealth data. Circles denote our six node types. Connections between circles denote our six edge types. The red line indicates the target ({\fontfamily{cmtt}\selectfont individual -  mental health}) edge type that we try to predict.}  \label{fig:1}
\end{figure}

\begin{figure}[h!]
\includegraphics[width=0.85\linewidth]{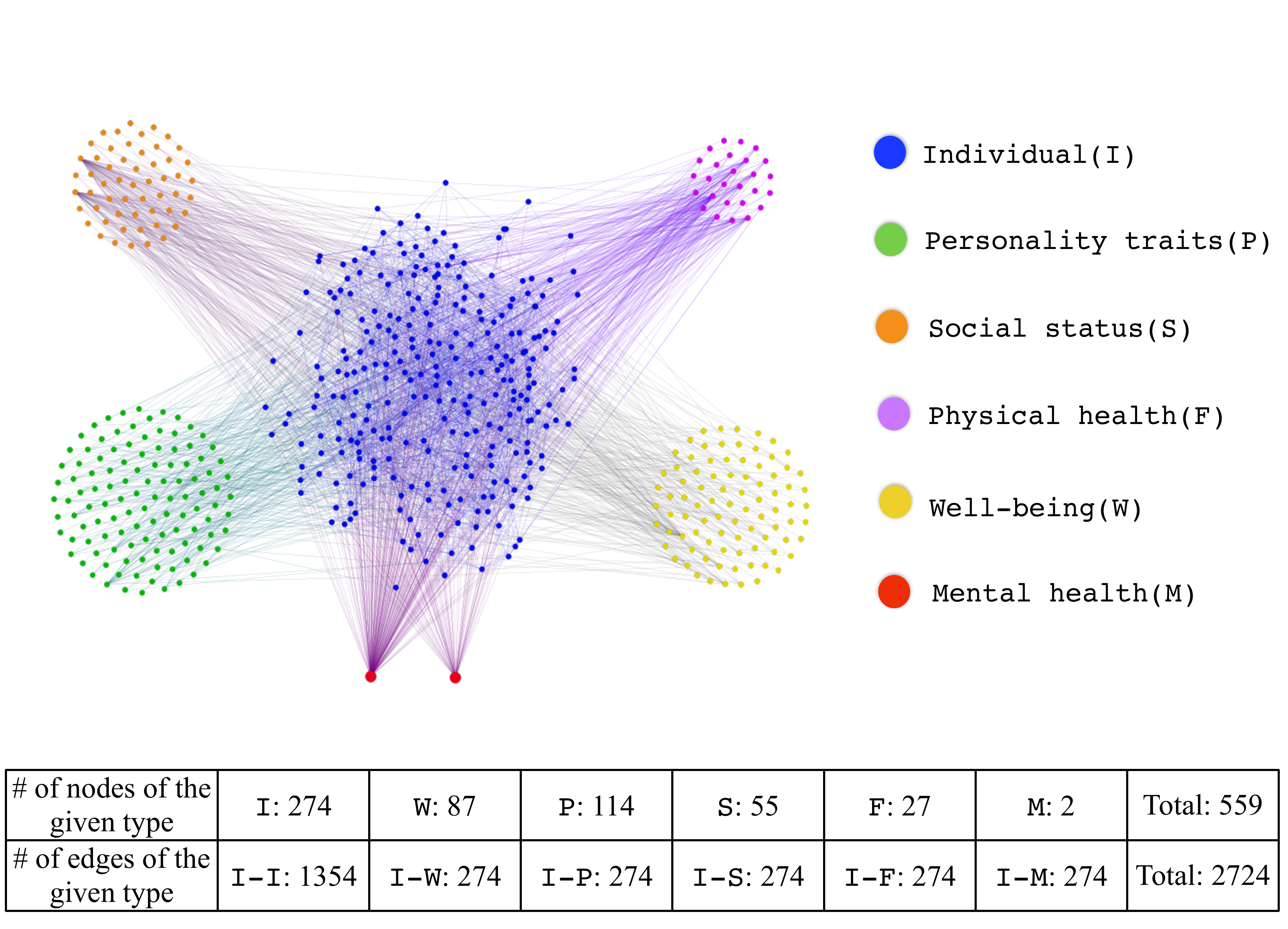}
\captionsetup{width=\textwidth}
\caption{Network visualization of the HIN constructed from the NetHealth data. Node colors represent the six node types from the network schema in Figure \ref{fig:1}. The table below the visualization summarizes the number of nodes or edges of each type and the total number of nodes or edges across all types. For detailed explanation behind the numbers of nodes and edges of the given type, see Section \ref{sect:network construction}. }  \label{fig:2}
\end{figure}

To integrate the different types of information, we model the NetHealth data as a \textit{heterogeneous information network (HIN)}, which is an effective tool to fuse information by considering multiple types of nodes and edges \cite{shi2017survey}. As a promising paradigm, HIN analysis has been applied to a variety of computational and applied tasks, such as recommendation systems in predicting movie ratings \cite{2019HINREC}, node classification in predicting authors or venues in academic networks \cite{dong2017metapath2vec}, clustering in visualization tasks \cite{Wang2014Ranking,sun2009rankclus}, link prediction in inferring future co-authorships  \cite{DBLP:conf/kdd/ZhangSHSC19}, and network alignment in predicting protein function from biological data \cite{gu2018homogeneous}. In our study, we apply HIN analysis to the task of mental health prediction. Figures \ref{fig:1} and \ref{fig:2} show the network schema and visualization of our HIN, respectively. Our constructed HIN consists of six node types ({\fontfamily{cmtt}\selectfont individual}, {\fontfamily{cmtt}\selectfont personality traits}, {\fontfamily{cmtt}\selectfont social status}, {\fontfamily{cmtt}\selectfont physical health}, {\fontfamily{cmtt}\selectfont well-being}, and {\fontfamily{cmtt}\selectfont mental health}). For the {\fontfamily{cmtt}\selectfont individual} node type, different nodes correspond to different student participants of the NetHealth study. For each of the other five node types ({\fontfamily{cmtt}\selectfont personality traits}, {\fontfamily{cmtt}\selectfont social status}, etc.), different nodes correspond to different personal characteristics of the given type, i.e., different values of the components listed under the given dimension in Figure \ref{fig:3}. For example, a node of the {\fontfamily{cmtt}\selectfont personality traits} type represents some combination of an individual's agreeableness, conscientiousness, extroversion, neuroticism, and openness. As an illustration, one combination, i.e., a node of the {\fontfamily{cmtt}\selectfont personality traits} type, may be low agreeableness - low conscientiousness - low extroversion - low neuroticism - low openness, and another combination, i.e., another node of the {\fontfamily{cmtt}\selectfont personality traits} type, may be high agreeableness - high conscientiousness - high extroversion - high neuroticism - high openness. See Section \ref{sect:network construction} for details. Clearly, there are multiple nodes of each type because there are multiple possible combinations of personal characteristics of the given type, and each combination corresponds to one node of that type. We connect the nodes of the network by forming six edge types. We construct the {\fontfamily{cmtt}\selectfont individual - individual} edge type by connecting two individuals if they communicate through SMS at least once during our study period. We construct edges of the other five types by connecting the {\fontfamily{cmtt}\selectfont individual} node type to each of the other five node types if the given individual has the given combination of personal characteristics of the given type. See Section \ref{sect:network construction} for details.

We construct the HIN in this way in order to be able to predict an individual's mental health state (likelihood of being depressed or anxious) by relying on the information about: 1) the individual's combinations of the four traits (personality traits, social status, physical health, and well-being), 2) the individual's position in their social network (i.e., which other individuals they are linked to), 3) combinations of the four traits of the individual's (direct or indirect) network neighbors, and 4) mental health states of the individual's (direct or indirect) network neighbors. For example, we argue that if an individual is linked (directly or indirectly) to many other individuals with whom she/he shares many combinations of the four traits, and if these individuals are depressed or anxious, then the individual in question is also likely to be depressed or anxious. So, our HIN offers an elegant and convenient yet important way to model influence between individuals.

More specifically, we have recognized that we can model the problem of mental health prediction from our HIN in a novel manner, as applying to the HIN a popular paradigm of a recommender system (RS). RS is widely used to suggest a personalized list of items (e.g., movies, books, or new friends) to individuals based on their preferences to help them find the most relevant items \cite{ricci2011introduction}. In other words, for an individual $i$,
based on the history of $i$'s behaviors (e.g., rating movies, liking contents on social media, or friendships with other individuals), an RS approach calculates a personalized ranking score on a set of new items (i.e., movies, contents, or friends, respectively) and suggests the top-ranked items to $i$. \cite{ricci2011introduction,vahedian2017multirelational}. In our case, items are the individuals' different mental health states, i.e., we predict what mental health state an individual is likely to have. 

Unlike their homogeneous counterparts that consider a single type of individual-item interaction (i.e., edge), HIN-based RS approaches take advantage of multiple node and edge types \cite{2019HINREC}. Multiple types of HIN-based RS approaches exist. 
One major approach category is based on matrix factorization. For example, DEDICOM, one of the earliest such approaches, is able to capture correlations between different types of nodes through matrix factorization. Such correlations are captured by including multiple edge types into the learning task at hand \cite{bader2007temporal}. RESCAL can be considered as a mathematically relaxed version of DEDICOM and was shown to perform better in recommendation tasks \cite{nickel2011three}. DMF further used novel objective functions compared to DEDICOM and RESCAL to learn latent feature parameters, which led to additional improvement of the prediction performance in recommendation tasks \cite{drumond2014optimizing}. 
Another major approach category  incorporates into the recommendation process the notion of a metapath, which follows a specific sequence of multiple types of nodes and edges of a HIN to capture additional HIN information for recommendation. For example, 
HeteRec learns metapath-based latent features (representations or embeddings)  of nodes (individuals and items) through a Bayesian ranking optimization technique \cite{DBLP:conf/wsdm/YuRSGSKNH14}.
Similarly, HERec learns node features through metapath-based random walks, but those are then transformed by fused functions and integrated into a matrix factorization framework with the goal of deriving more informative node features for recommendation \cite{2019HINREC}.
Unlike the above methods that cannot deal with weighted graphs (i.e., edges), SemRec aimed to account for edge weights, and it did so via weighted metapaths in RS to generate better recommendations \cite{Shi2015Semantic}.  
Different from the many traditional RS methods that are based on matrix factorization, MCRec used a deep neural network with a co-attention mechanism to leverage the metapath-based context for recommendation \cite{2018leveragingmeta}. 
Unlike the above methods that did not consider social (i.e., individual-individual) interaction data, DSR accounted for edges between individuals and used an extended social similarity regularization, which imposed constraints on both similar and dissimilar nodes in its RS framework \cite{Jing2016Recommendation}. 
Similarly, SimMF integrated both social interactions and attribution information of individuals and items to make recommendations; it did so via metapath similarity measures  \cite{Shi2015Integrating}.

Of all HIN-based RS methods mentioned above, we apply  four prominent or recent ones, namely DMF \cite{drumond2014optimizing}, HERec \cite{2019HINREC}, RESCAL \cite{nickel2011three}, and DEDICOM \cite{bader2007temporal}, to the HIN to predict edges between nodes of the {\fontfamily{cmtt}\selectfont individual} type and nodes of the {\fontfamily{cmtt}\selectfont mental health} type, denoted as \textit{target edges} in our HIN (Figure \ref{fig:1}). 

In addition, to evaluate the power of RS, we model the problem of mental health prediction in an alternative manner, as node classification (NC) in our HIN. NC predicts labels of nodes based on their features that characterize positions of nodes in a network \cite{bilgic2007combining}. In our case, node labels are individuals' different mental health states.
In our study, we extract four prominent features of individuals from our HIN: graphlets \cite{milenkovic2008uncovering}, colored graphlets \cite{gu2018homogeneous}, DeepWalk \cite{perozzi2014deepwalk}, and Metapath2vec++ \cite{dong2017metapath2vec}. Then, we put network features into a logistic regression classifier to make predictions, which is a common approach in NC \cite{goyal2018graph}. 
Note that we choose these four features for the following reasons. The first two are based on
homogeneous and heterogeneous graphlets, respectively, where graphlets are small subgraphs, i.e., Lego-like building blocks, of complex networks \cite{milenkovic2008uncovering,gu2018homogeneous}. The graphlet-based features, which intuitively capture how many subgraphs of each type a node participates in, were extensively demonstrated  to be powerful measures of the network position of a node in numerous  tasks \cite{bookchap2019}. The last two are prominent homogeneous and heterogeneous network embedding methods, respectively.  Network embedding has received significant attention in the last several years, owing to of its ability to \emph{automatically} learn low-dimensional latent node features that likely preserve the original high-dimensional network structure \cite{perozzi2014deepwalk,dong2017metapath2vec,DBLP:conf/kdd/ZhangSHSC19,DBLP:conf/kdd/ShiZGZ018}.

To evaluate the prediction performance of the four RS and four NC network methods, we compare them against a logistic regression classifier using the same NetHealth data in a \textit{non-network} fashion \cite{mohr2017personal}, and against a random guess method \cite{gatsoulis2016learning}. \textbf{To our knowledge, this is the first study to use smartphone, wearable sensor, and survey data in an HIN model and use RS or NC on the HIN to predict individuals' mental health conditions. }   


Our findings are as follows.
\begin{itemize}

\vspace{-1mm}

\item For both depression and anxiety prediction, among all network methods, DMF (an RS method) makes the most accurate predictions. Compared to the random guess method, for both depression and anxiety, all RS and NC methods except for RESCAL and DEDICOM are significantly more accurate. Compared to the non-network method, for depression, DMF and DeepWalk are significantly more accurate, graphlets, colored graphlets, and Metapath2vec++ are marginally more accurate, HERec is comparable, and RESCAL and DEDICOM are significantly less accurate; and for anxiety, DMF is significantly more accurate, DeepWalk is marginally more accurate, HERec, graphlets, colored graphlets, and Metapath2vec++ are marginally less accurate, and RESCAL and DEDICOM are significantly less accurate. Our results indicate that (at the minimum) the best of the network methods outperforms the random guess method and the non-network method. In addition, the best RS method outperforms all NC methods. This confirms the power of networks and RS in particular in predicting mental health.

\item We explore whether the different types of methods (RS, NC, non-network) identify different sets of depressed/anxious individuals. If this is true, we might be able to make more accurate predictions by combining the different method types. Specifically, we study overlaps between depressed/anxious individuals \textit{correctly} predicted by three representative methods: the most accurate RS method (DMF), the most accurate NC method (DeepWalk), and the non-network method. For depression, we find that DMF's and DeepWalk's predictions significantly overlap. Additionally, DeepWalk's and the non-network method's predictions also significantly overlap. But DMF's and the non-network method's predictions do not significantly overlap. For anxiety prediction, we observe similar results. The reasons for these observations could be: 1) DMF and DeepWalk are both network methods and differ only in one aspect - they are different types of methods (RS versus NC); 2) DeepWalk and the non-network method both use the logistic regression to make predictions and differ only in one aspect - DeepWalk uses network features while the non-network method does not; and 3) DMF differs from the non-network method in two ways: the former is a network method and it does not use logistic regression to make predictions, while the latter is a non-network method that uses logistic regression. In other words, it could be that the more similar the two approaches are in terms of their methodologies, the more similar their predictions. Our results suggest that because of the non-significant overlap between DMF's and the non-network method's predictions, by combining the two methods, we might be able to make more accurate predictions. Exploring this is beyond the scope of this paper and is the subject of our future work.

\item In our prediction approaches mentioned above, we have integrated all types of information about individuals, represented by the different edge types in the HIN, into our (RS and NC) network methods to make mental health predictions. To study whether some edge types might be more informative (i.e., have more predictive power) than others, we investigate the effect of using all possible combinations of edge types (including all edge types) on the prediction performance. For this analysis, we focus only on DMF as the most accurate of all analyzed methods in the above evaluations. We consider a series of DMF instances that correspond to all possible combinations of all five non-target edge types from our HIN (the target edge type needs to be always included by default). For example, on the one extreme, we consider each of the five non-target edge types alone, and on the other extreme, we consider all five non-target edge types combined (which has been the case in the above evaluations). An example of an edge type combination in-between these two extremes is the combination of the {\fontfamily{cmtt}\selectfont individual - individual}, {\fontfamily{cmtt}\selectfont individual - physical health}, and {\fontfamily{cmtt}\selectfont individual - personality traits} edge types. We find that for depression, the combination of the {\fontfamily{cmtt}\selectfont individual -  physical health} and {\fontfamily{cmtt}\selectfont individual - well-being} edge types is significantly more accurate than the rest of the combinations, including the combination of all edge types. For anxiety, the combination of the {\fontfamily{cmtt}\selectfont individual - personality traits} and {\fontfamily{cmtt}\selectfont individual - well-being} edge types is marginally more accurate than six other edge type combinations, including the combination of all edge types, and is significantly more accurate than the rest of the combinations. Our results indicate that we can make more accurate predictions using some subset of edge types than using all edge types. Exploring the reasons behind this observation is beyond the scope of this paper and is the subject of our future work.

\end{itemize}

\section{Methods}
\subsection{HIN construction} \label{sect:network construction}

\textbf{Specific data used in this paper.} \hspace{1mm} The data used in this paper come from the NetHealth study, an institutional review board-approved effort that collected smartphone, Fitbit, and survey data from approximately 700 undergraduate student volunteers at the University of Notre Dame, who entered the study as freshmen in August 2015 \cite{purta2016experiences, faust2017exploring}. Smartphone data was collected via a monitoring application (CIMON) installed on individuals' (student participants') smartphones that periodically synchronizes SMS logs \cite{hossain2016challenges}. Notably, we did not collect information about the content of the messages. Instead, we collected information on who communicates with whom and when. Health behavioral data was collected through Fitbit Charge HR devices. In this study, we consider two specific Fitbit metrics, individuals' step counts and their sleep duration. We conducted periodic surveys that asked questions about individuals' demographics, personality traits, mental health, and other subjects. From surveys, we extract 15 characteristics that can be divided into five categories: 1) personality traits: agreeableness, conscientiousness, extraversion, neuroticism, and openness, as measured by the Big-Five personality test \cite{goldberg1990alternative}; 2) social status: gender, race, religion, and parents' educations; 3) physical health: sleep quality measured by the Pittsburgh Sleep Quality Index (PSQI) \cite{buysse1991quantification}; 4) well-being: body image, happiness, health, loneliness \cite{cramer2000abbreviated}, and self-esteem \cite{robins2001measuring}; and 5) mental health: depression measured by the Center for Epidemiological Studies Depression Scale (CES-D) test \cite{lewinsohn1997center} and anxiety measured by the State-Trait Anxiety Inventory (STAI) test \cite{spielberger2010state}.

\vspace{0.2cm}
\noindent \textbf{Selection of the study time period and pool of individuals.} \hspace{1mm} In this paper, we select one year from August 2015 to August 2016 as the study period since our previous study showed that the majority of individuals are actively involved in the NetHealth study during this period but not during other periods \cite{liu2018network}. This period covers 52 weeks, of which 31 are school weeks and 21 are break weeks. In this study, we focus only on school weeks because our previous study showed that school weeks have meaningful social network structures, while break weeks do not \cite{liu2018network}. The NetHealth data contain SMS logs of 615 iPhone users and 96 Android users. In this study, we only consider iPhone data since we encountered problems with Android data consistency. Finally, we keep only SMS data where both the sender and receiver are among the 615 iPhone users, i.e., the participants of our study. In other words, we discard SMS data where the sender or receiver is not a participant of the NetHealth study (e.g., is a participant's family member or friend).

We use the following three criteria established in our previous study \cite{liu2018network} to select individuals who have high-quality data:
\begin{enumerate}
    \item Students' first SMS activity date was before or on the start date of our considered study period and their last SMS activity was on or after the end date of our considered study period, i.e., students actively sent or received SMSs during the one-year study period.
    
    \item Students have high-quality Fitbit data, i.e., their Fitbit data are valid. By ``valid'', we mean that a student wore the Fitbit at least 80 percent of the time in more than half of the days during the one-year study period.
    \item Students completed the survey taken in July 2015, which was conducted before students entered the school, and the survey taken in January 2016, which was conducted during our study period. The reasons why we need both surveys are discussed below.
    
\end{enumerate}

274 out of the 615 individuals satisfy the three criteria and thus form the final pool of individuals considered in this study.

\vspace{0.2cm}
\noindent \textbf{The HIN.} \hspace{1mm} One of the key contributions of this study is to establish an HIN using the NetHealth data. An HIN is a network consisting of multiple types of nodes and edges \cite{sun2009rankclus}. In this study, we construct an HIN consisting of six node types and six edge types (listed below). Formally, the HIN can be represented as $G=(V,E)$, where $V = \bigcup_{i}V_i$, $i\in\{1,2,3,4,5,6\}$, i.e., $V$ is the union of all nodes over all six node types, and $E = \bigcup_{j}E_ji$, $j\in\{1,2,3,4,5,6\}$ i.e., $E$ is the union of all edges over all six edge types.

The node types include {\fontfamily{cmtt}\selectfont individual}, {\fontfamily{cmtt}\selectfont personality traits}, {\fontfamily{cmtt}\selectfont physical health}, {\fontfamily{cmtt}\selectfont social status}, {\fontfamily{cmtt}\selectfont well-being}, and {\fontfamily{cmtt}\selectfont mental health}, as follows (also, see Figures \ref{fig:1} and \ref{fig:2}).

\begin{itemize}
    \item The {\fontfamily{cmtt}\selectfont individual} node type represents students who participate in the NetHealth study. There are 274 nodes of the {\fontfamily{cmtt}\selectfont individual} node type.
    
    \item The {\fontfamily{cmtt}\selectfont personality traits} node type represents a combination of five personality characteristics including agreeableness, conscientiousness, extroversion, neuroticism, and openness, extracted from the survey taken in January 2016. We use data from this survey because it is the only one conducted during our study period. The distributions of personality characteristics scores are shown in Supplementary Figure S1. We combine these five characteristics because they belong to the same type of the individuals' information, as also suggested by domain experts who have conducted the NetHealth study. The individuals' personality characteristics scores vary from person to person. For example, the individuals' agreeableness scores could be 2.1, 2.2, 3.4, 4.2, 4.8, etc. Combining such absolute score values across the different personality characteristics and over all individuals would generate a huge number of possible combinations. To address this in an elegant way, for each personality characteristic, we divide the individuals, i.e., their scores, into three groups, as is typically done \cite{sano2015recognizing}: the high-score group (the top 25 percent of the scores), the medium-score group (the middle 50 percent of the scores), and the low-score group (the lowest 25 percent of the scores). Then, we form nodes of the {\fontfamily{cmtt}\selectfont personality traits} type by combining the group labels of the five personality characteristics. 
    For example, a node of the {\fontfamily{cmtt}\selectfont personality traits} type may be low agreeableness - medium conscientiousness - medium extroversion - medium neuroticism - high openness. The number of all possible combinations of the five personality characteristics is $3^5=243$. Note that some possible combinations are not present in the NetHealth data. So, the number of combinations that correspond to scores of at least one individual in our data, i.e., the number of nodes of the {\fontfamily{cmtt}\selectfont personality traits} type, is $114$.

    \item The {\fontfamily{cmtt}\selectfont social status} node type represents a combination of four social status characteristics including gender, race, religion, and parents' education, extracted from the survey taken in July 2015. We use social status data from this survey because it is the only one containing the individuals' social status information. (This is the reason why we need both surveys in the third criterion when choosing which pool of individuals to consider.) The distributions of the 274 individuals' social status scores are shown in Supplementary Figure S2. For example, a node of the {\fontfamily{cmtt}\selectfont social status} type may be female - white - Catholic - both parents received bachelor's degrees. Given two possible genders, five possible races, four possible religions, and three possible options for parents' education, the number of all possible combinations of the four social status characteristics is $2 \times 5 \times 4 \times 3 = 120$. The number of combinations that correspond to scores of at least one individual in our data, i.e., the number of nodes of the {\fontfamily{cmtt}\selectfont social status} type, is $55$.
     
    \item The {\fontfamily{cmtt}\selectfont physical health} node type represents a combination of three physical health characteristics including sleep quality, average sleep duration, and average step counts. We obtain the individuals' sleep quality scores from the survey taken in January 2016 and their average sleep duration and average step counts during the study period from the Fitbit data. The distributions of the scores are shown in Supplementary Figure S3. To combine sleep quality, average sleep duration, and average step counts, we divide scores of each physical health characteristic into three groups: the high-score group (the top 25 percent of the scores), the medium-score group (the middle 50 percent of the scores), and the low-score group (the lowest 25 percent of the scores). Then, we form nodes of the {\fontfamily{cmtt}\selectfont physical health} type by combining the group labels of the three physical health characteristics. For example, a node of the {\fontfamily{cmtt}\selectfont physical health} type may be low sleep quality - medium average sleep duration - high step counts. The number of all possible combinations of the three physical health characteristics is $3^3=27$. The number of combinations that correspond to scores of at least one individual in our data, i.e., the number of nodes of the {\fontfamily{cmtt}\selectfont physical health} type, is $27$.

    \item The {\fontfamily{cmtt}\selectfont well-being} node type represents a combination of five well-being characteristics including body image, happiness, health, loneliness, and self-esteem, extracted from the survey taken in January 2016. The distributions of the scores are shown in Supplementary Figure S4. To combine the five well-being characteristics, we divide scores of each characteristic into three groups: the high-score group (the top 25 percent of the scores), the medium-score group (the middle 50 percent of the scores), and the low-score group (the lowest 25 percent of the scores). Then, we form nodes of the {\fontfamily{cmtt}\selectfont well-being} type by combining the group labels of the five well-being characteristics. For example, a node of the {\fontfamily{cmtt}\selectfont well-being} type may be low body image - medium happiness - medium health - medium loneliness - high self-esteem. The number of all possible combinations of the five well-being characteristics is $3^5=243$. The number of combinations that correspond to scores of at least one individual in our data, i.e., the number of nodes of the {\fontfamily{cmtt}\selectfont well-being} type, is $87$.

    \item The {\fontfamily{cmtt}\selectfont mental health} node type represents a mental health condition: either depression or anxiety. (Technically, we analyze each of depression and anxiety via its respective HIN.) For depression, we use one node to denote having ``depressed'' condition, and another node to denote having ``non-depressed'' condition. Similarly, for anxiety, we use one node to denote having ``anxious'' condition, and another node to denote having ``non-anxious'' condition. 
    The individuals' depression and anxiety scores are collected from the survey taken in January 2016. We consider individuals whose depression scores are above 15 as depressed, which is suggested in the literature \cite{kohout1993two,jackson2010race,rosenquist2011social}. Out of the 274 individuals, 67 individuals (24.5\%) are depressed according to this criterion. We consider individuals whose anxiety scores are above 40 as anxious, which is suggested in the literature \cite{grant2008maternal,julian2011measures,faisal2007prevalence}. Out of the 274 individuals, 106 individuals (38.7\%) are anxious according to this criterion. 
    
\end{itemize}

The six edge types in our HIN include {\fontfamily{cmtt}\selectfont individual -  individual}, 
{\fontfamily{cmtt}\selectfont individual -  personality traits}, 
{\fontfamily{cmtt}\selectfont individual - social status}, 
{\fontfamily{cmtt}\selectfont individual - physical health}, 
{\fontfamily{cmtt}\selectfont individual -  well-being}, and
{\fontfamily{cmtt}\selectfont individual -  mental health}, as follows (also, see Figures \ref{fig:1} and \ref{fig:2}).  

\begin{itemize}
\item The {\fontfamily{cmtt}\selectfont individual - individual} edge type is constructed by connecting two individuals if they have communicated through SMS at least once within our study period. This results in 1354 edges between our 274 individuals. We assign a weight to each edge, corresponding to the total number of SMSs exchanged between two given individuals within the study period. 

\item The remaining five edge types are constructed by connecting the {\fontfamily{cmtt}\selectfont individual} node type to the other five node types ({\fontfamily{cmtt}\selectfont personality traits}, {\fontfamily{cmtt}\selectfont physical health}, {\fontfamily{cmtt}\selectfont social status}, {\fontfamily{cmtt}\selectfont well-being}, and {\fontfamily{cmtt}\selectfont mental health}). Here, an edge is formed if the given individual has the given combination of personal characteristics of the given type. For example, if an individual has low agreeableness - low conscientiousness - low extroversion- low neuroticism - low openness, we connect the individual to the node of the {\fontfamily{cmtt}\selectfont personality traits} type representing such combination. In our data, an individual can only have one combination of personal characteristics of a given node type (e.g., for the {\fontfamily{cmtt}\selectfont personality traits} node type, if an individual is linked to a node corresponding to low agreeableness - low conscientiousness - low extroversion - low neuroticism - low openness, he/she cannot be linked to a node corresponding to high agreeableness - high conscientiousness - high extroversion - high neuroticism - high openness or to any other combination). Therefore, an individual is connected to exactly one node of each of the five node types. Thus, for each of the five edge types, the number of edges is the same as the number of the individuals, i.e., 274. The {\fontfamily{cmtt}\selectfont individual - mental health} edge type is considered as the target edge type since this is what we try to predict.

\end{itemize}


\subsection{Network methods that we adapt to the task of predicting mental health conditions} \label{sec:methods}

We model the problem of predicting individuals' mental health conditions as RS and NC. 

\vspace{0.2cm}
\noindent \textbf{RS methods.} \hspace{1mm}
RS is widely used to suggest a personalized list of items (e.g., movies, books, or new friends) to individuals based on their preferences, in order to help them find the most relevant items \cite{ricci2011introduction}. In this study, we use four RS methods: DMF \cite{drumond2014optimizing}, RESCAL \cite{nickel2011three}, DEDICOM \cite{bader2007temporal}, and HERec \cite{2019HINREC}. 

The first three methods are of the multi-relational matrix factorization (MRMF) type \cite{krohn2012multi}. In MRMF, the target edge type is to be predicted and the remaining types of edges as used as side information. For example, if the task is to recommend movies to individuals, the individual - movie edge type is used as the target edge type and other types of edges such as movie - genre, movie - actor, etc. are used as side information. An MRMF model is trained using a proportion of target edges as well as side information, and then the model is used to predict the rest of the target edges. MRMF computes latent features of all types of nodes, which are used to map individuals to items (the edge type to be predicted) that the individuals will prefer. However, instead of using only one edge type, as in standard factorization schemes \cite{harshman1994parafac}, MRMF allows for the creation of additional latent features based on other side information (other edge types). By operating on latent features between a pair of nodes, we can obtain a value denoting the probability that the given node pair has an edge. MRMF optimizes the loss function to minimize the difference between probability values of all node pairs (represented by a matrix) and the adjacency matrix representing real edges. The three MRMF methods that we consider use different loss functions (that is, they differ in how they mathematically define the difference between probability values of all node pairs and the adjacency matrix) and different optimization methods (that is, they differ in how they minimize their respective loss functions) \cite{vahedian2015network,vahedian2016meta}.  We use all three MRMF-based RS approaches to predict target edges (i.e., edges of the {\fontfamily{cmtt}\selectfont individual -  mental health} type). 
For more details on the three MRMF RS methods, see Supplementary Section S1.

As a different approach type, HERec works as follows. First, it learns latent features of nodes from the HIN through metapath-based random walks. For this step, to mimic as closely as possible the case studies in the HERec paper \cite{2019HINREC}, i.e., to give HERec the best-case advantage, we consider all possible metapath types from  the {\fontfamily{cmtt}\selectfont individual} node type to  the {\fontfamily{cmtt}\selectfont individual} node type via each of the other five considered node types. That is, we consider the following five types of metapaths: {\fontfamily{cmtt}\selectfont individual - well-being - individual}, {\fontfamily{cmtt}\selectfont individual - personality traits - individual}, {\fontfamily{cmtt}\selectfont individual - social status - individual}, {\fontfamily{cmtt}\selectfont individual - physical health - individual}, and {\fontfamily{cmtt}\selectfont individual - mental health - individual}. Second, to derive more effective (i.e., better-informed) node representations, HERec uses fused functions to transform the latent node features from the the previous step into a more suitable form for recommendations. For this second step,  we consider the personalized non-linear fusion function, which was shown to yield the best performance in the recommendation task among three  functions considered in the HERec paper \cite{2019HINREC}. Third, HERec blends the transformed features into a matrix factorization framework to predict rating scores of individuals. For this step, we model the problem of rating prediction as the task of predicting {\fontfamily{cmtt}\selectfont individual - mental health} links (i.e., target edges). Namely, we use HERec to predict rating scores of all individuals,  sort the predicted scores from highest to lowest, and select a cutoff where individuals with scores above the cutoff will be predicted as depressed (anxious) and those with scores below the cutoff as non-depressed (non-anxious). Note that HERec has the node feature dimension as input parameter. The method uses the default value of 30 for this parameter. To give  HERec the best-case advantage, we evaluated additional values of this parameter, namely 10 and 20, i.e., lower-dimensional features than used in the HERec paper, because our data has fewer nodes than the data from the HERec study. However, we found that reducing the dimension decreased HERec's perfrmance. So, we ended up reporting results for the default setting, as this option yielded the best accuracy.

When we predict target edges of the {\fontfamily{cmtt}\selectfont individual} -  {\fontfamily{cmtt}\selectfont mental health} type using the four RS methods discussed above, we randomly divide the individuals into 80\% training individuals and 20\% testing individuals. This step is conducted using stratified sampling to ensure that the split data maintain the percentage of the individuals who are depressed (or anxious). 
We use target edges of training individuals as well as the other types of edges to train an RS model, and then use the model to predict target edges of testing individuals. In other words, we hide a proportion of target edges and use the remaining edges to train the model, and then use the model to predict the hidden target edges. Formally, given the HIN $G=(V,E)$, let $I \in V$ and $M \in V$ denote the sets of all individuals and all mental health conditions, respectively. Then, the goal is to predict  \{$i_{test}$--$m$\}, i.e., the target edges of the testing individuals, where $i_{test}\in I$ and $m\in M$. To predict an individual as depressed or non-depressed, we calculate probability values of the two target edges that connect the individual to the node representing ``depressed'' condition and the node representing ``non-depressed'' condition, respectively. We predict an individual as anxious or non-anxious in the same manner. Of the two target edges, we select the one with the higher probability value as the final prediction.

\vspace{0.2cm}
\noindent \textbf{NC methods.} \hspace{1mm}
We model the problem of mental health prediction in an alternative manner, as NC that uses the structural information encoded in the network to predict nodes' labels. For example, in a social network consisting of individuals with known labels being smoker or non-smoker and individuals with unknown labels, the NC method predicts the latter as smokers or non-smokers based on their network features, i.e., their network ``relationships'' with the former \cite{bilgic2007combining}. Network features characterize properties of nodes in a network, such as nodes' network positions and their neighborhood information. In order to make mental health predictions using NC methods, our first step is to extract the individuals' network features from our constructed HIN. We learn the individuals' network features based on all edge types except for the target edge type. Instead, we use the target edge type to label the individuals according to their mental health states, for the purpose of NC training and testing (see below). For example, if an individual connects to a node representing the ``depressed'' condition, we assign a ``depressed'' label to the individual. We extract four features for the individuals: graphlets \cite{milenkovic2008uncovering}, colored graphlets \cite{gu2018homogeneous}, DeepWalk \cite{perozzi2014deepwalk}, and Metapath2vec++ \cite{dong2017metapath2vec}. Among these four, graphlets and DeepWalk learn features of nodes from a homogeneous network containing a single node type and a single edge type; in other words, when applied to our HIN, they ignore the different node and edge types and consider the network as homogeneous. Colored graphlets and Metapath2vec++ learn features of nodes from a heterogeneous network consisting of multiple types of nodes and edges. We use the two homogeneous network methods because 1) these methods are widely used to extract features of nodes from a network and 2) there is a limited number of methods that extract nodes' features from a heterogeneous network. In more detail, graphlets are small connected non-isomorphic induced subgraphs of a network \cite{milenkovic2008uncovering}. Based on the notion of graphlets, the graphlet degree vector (GDV) is a network feature that summarizes a node's extended network neighborhood. In this study, we use both homogeneous graphlets \cite{milenkovic2008uncovering} and heterogeneous graphlets \cite{gu2018homogeneous} to calculate nodes' GDVs. The DeepWalk method \cite{perozzi2014deepwalk} uses random walks to generate node sequences in a homogeneous network. The node sequences are put into a skip-gram language model to learn network features. The Metapath2vec++ method \cite{dong2017metapath2vec} formalizes metapath-based random walks which are restricted to only transitions between certain types of nodes. Then, Metapath2vec++ leverages a heterogeneous skip-gram model to learn nodes' features vectors from the metapath-based random walks. For more details on the four NC methods, see Supplementary Section S1.

In NC, we use the same training and testing individuals that we have used in RS in order to compare the different types of network methods in an unbiased way. After learning the individuals' network features, the second step in NC is to train a classifier (e.g., logistic regression, random forest, or support vector machine) based on training individuals' network features and node labels and make predictions on testing individuals' node labels. Node labels in NC and target edges in RS are conceptually equivalent. We model target edges as node labels in NC and this is why we do not include target edges when learning network features of individuals. In this study, we use logistic regression as a proof-of-concept classifier, because this particular classifier is often used to predict individuals’ mental health conditions \cite{mohr2017personal,lin2016association,reece2017instagram}. The output of a logistic regression classifier is a set of probability scores, one score per individual. For each logistic regression classifier used in this study (one classifier per network feature), we choose a probability cutoff where individuals with probability scores above the cutoff value will be classified as depressed/anxious, and those with probability scores below the cutoff value will be classified as non-depressed/non-anxious. For each logistic regression classifier used in this study, we choose a cutoff value such that the proportion of the individuals who are predicted as depressed/anxious is equal to the proportion of the individuals who are actually depressed/anxious, as is typically done \cite{freeman2008comparison}. 

RS and NC are mathematically different since they work in different ways as discussed above. But ultimately, their inputs and outputs are the same, which makes them directly comparable. 

\vspace{0.2cm}
\noindent \textbf{Comparison of RS and NC with a non-network method and a random guess method.} \hspace{1mm}
To investigate the effectiveness of the RC and NC network methods, we compare the network methods to each other as well as to a non-network method \cite{mohr2017personal} and a random guess method \cite{gatsoulis2016learning}. For the non-network method and the random guess method, we use the same training and testing individuals that we have used in RS and NC. 

We implement the non-network method as follows, making it as fairly comparable to our RS and NC network methods as possible. \textbf{The non-network method is also a logistic regression classifier that utilizes the same NetHealth data as our network methods do but in a \textit{non-network} fashion.} Specifically, the non-network method uses the 15 characteristics collected from the surveys (Section \ref{sect:network construction}), average step counts and average sleep duration collected from Fitbits, and numbers of SMS messages sent or received during the study period collected from smartphones. For each of these data types, we divide individuals into three groups: the high-score group, the medium-score group, and the low-score group, which is the same approach that we have used when forming nodes of our HIN (Section \ref{sect:network construction}). We do this to ensure that the step of dividing individuals to groups is consistent and is thus not a factor that can account for differences in the results.

The random guess method works as follows. For example, in the depression prediction task, suppose that $x\%$ of the individuals are depressed and $(1-x)\%$ of the individuals are non-depressed. To make predictions, the random guess method randomly chooses $x\%$ of all individuals and predicts them as depressed, and it predicts the remaining individuals as non-depressed.

\subsection{Evaluation methodology} \label{sec:eval}
We use 5-fold cross-validation to evaluate the performance of the four RS and four NC methods, the non-network method, and the random guess method. We divide all individuals into five equal-sized subsets, such that each subset contains the same proportion of depressed/anxious individuals as present in the considered pool of individuals. We use one of the subsets as the testing set and the union of the other four subsets as the training set. We repeat this process five times until every subset has served as the testing set. We calculate the average and standard deviation of evaluation measures (defined below) over the five runs. Moreover, we make a prediction about an individual when the individual is part of the testing set. This way, we are able to predict mental health conditions for all individuals through the cross-validation process.      

Given a prediction for an individual, there are four outcomes \cite{fawcett2006introduction}. Taking depression as an example, a true positive (TP) represents an individual who is depressed and is also predicted as depressed. A false negative (FN) represents an individual who is depressed but is predicted as non-depressed. A true negative (TN) represents an individual who is non-depressed and is also predicted as non-depressed. A false positive (FP) represents an individual who is non-depressed but is predicted as depressed.

In our study, we consider the following four evaluation measures:
\begin{enumerate}
    \item $Precision = {TP \over (TP+FP)}$: of all predictions, how many are correctly predicted as depressed.
    \item $Recall = {TP \over (TP+FN)}$: of all depressed people, how many are correctly predicted as depressed. 
    \item $F1 \,score = {2TP \over (2TP+FP+FN)}$: the harmonic mean of precision and recall.
    \item $Accuracy = {(TP+TN) \over (TP+TN+FP+FN)}$: of all predictions, how many are correctly predicted (as either depressed or non-depressed).
\end{enumerate}

To statistically compare the performance of the different methods when predicting the individuals' mental health conditions, we use the Wilcoxon signed-rank test, which is a non-parametric test used to compare paired samples to assess whether their distributions differ \cite{rey2011wilcoxon}. Since for each method we compare its prediction performance against the rest of the methods that we evaluate, we adjust \textit{p}-values via false discovery rate (FDR) estimation to account for multiple test correction \cite{noble2009does}. In this study, we use the adjusted \textit{p}-value threshold of 0.05.

We explore whether the most accurate RS method (DMF), the most accurate NC method (DeepWalk), and the non-network method identify different sets of individuals as depressed/anxious. If this is true, we might be able to make more accurate predictions by combining the different method types. For each pair of the three methods, we examine whether sets of depressed/anxious individuals correctly predicted by the two methods are significantly overlapping through a hypergeometric test. Suppose that $S$ is the set of all depressed individuals, $A$ is the set of depressed individuals correctly predicted by one of the two methods, $B$ is the set of depressed individuals correctly predicted by the other method, and $O$ is the overlap between $A$ and $B$. Then, the $p$-value (i.e., the probability of obtaining the overlap of size |O| or greater) is:
\begin{equation} 
P (X \geq |O|) = 1 - \sum_{i=0}^{|O|-1}\frac{{S \choose i}{|S|-|A| \choose |B|-i}}{{|S| \choose |B|}}
\end{equation} 
We say that the overlap is significant if its $p$-value is $<$ 0.05.


\section{RESULTS}\label{sect:results}

In this study, we investigate three research questions: 
\begin{itemize}

\item Q1: How do RS, NC, non-network, and random approaches compare to each other in terms of accuracy of predicting individuals' mental health?

\item Q2: Do the most accurate RS method, the most accurate NC method, and the non-network method identify different sets of anxious/depressed individuals, i.e., are they complementary to each other?

\item Q3: What is the impact of using different types of information about the individuals, represented by different edge types in the HIN, on the performance of mental health prediction?

\end{itemize}

\noindent \textbf{Q1: How do RS, NC, non-network, and random approaches  compare to each other in terms of accuracy of predicting individuals' mental health?}

\begin{figure}[h!]
    \centering
    \begin{subfigure}[b]{0.45\textwidth}
    \includegraphics[width=\linewidth]{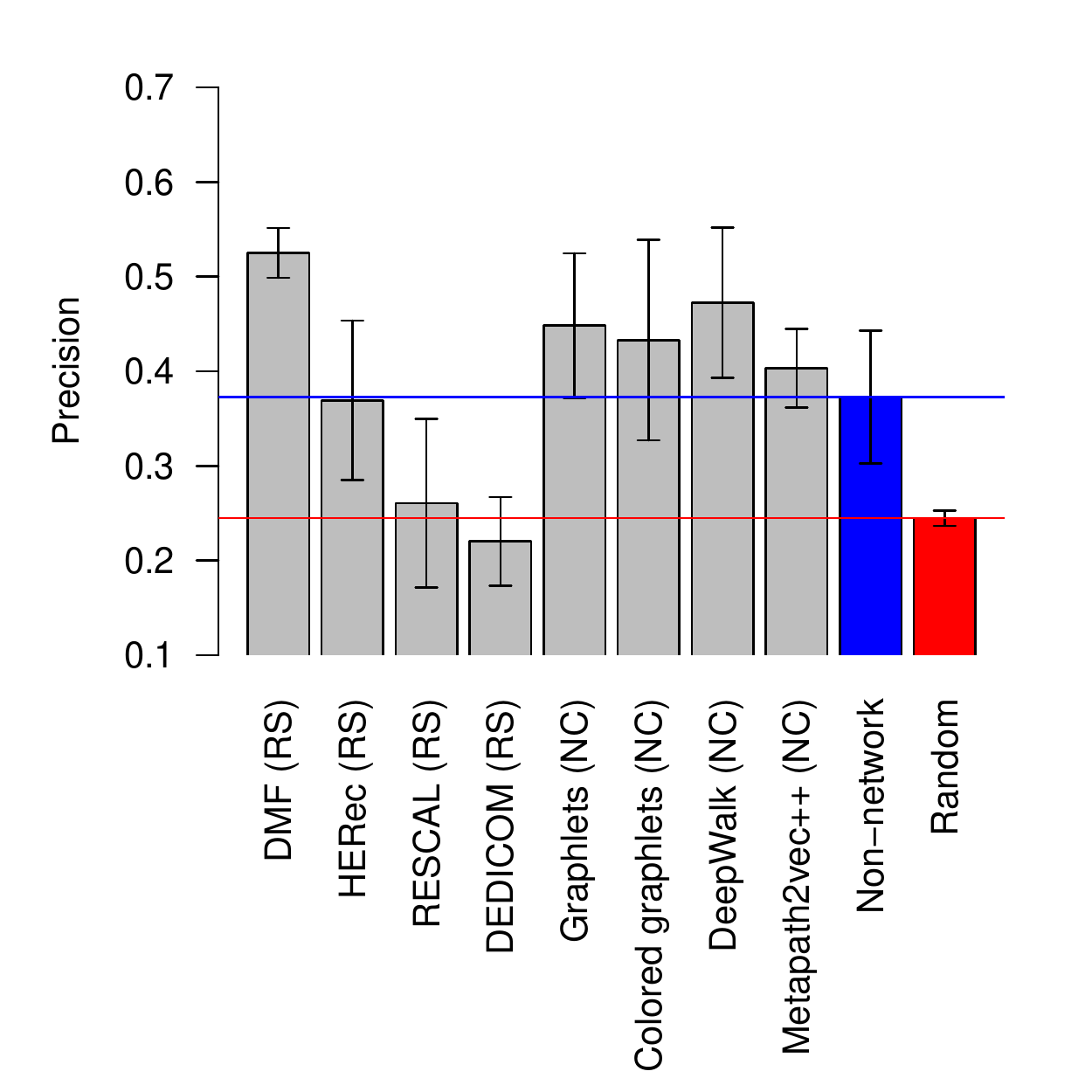}
    \captionsetup{width=\textwidth}
\caption{Depression}
\label{fig:4a}
\end{subfigure}
\hfill
\begin{subfigure}[b]{0.45\textwidth}
    \centering
    \includegraphics[width=\linewidth]{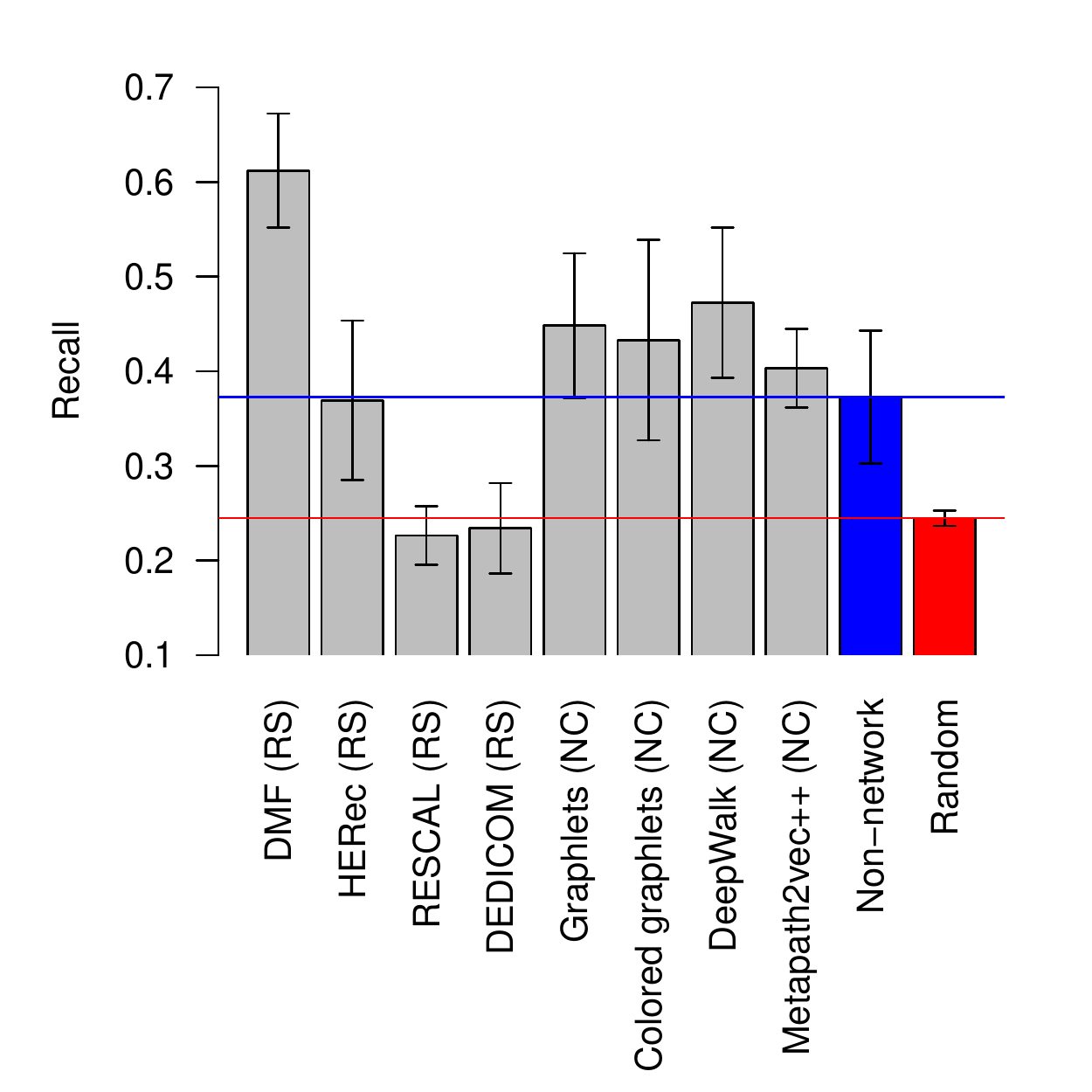}
    \captionsetup{width=\textwidth}
\caption{Depression}
\label{fig:4b}
\end{subfigure} 

\begin{subfigure}[b]{0.45\textwidth}
    \includegraphics[width=\linewidth]{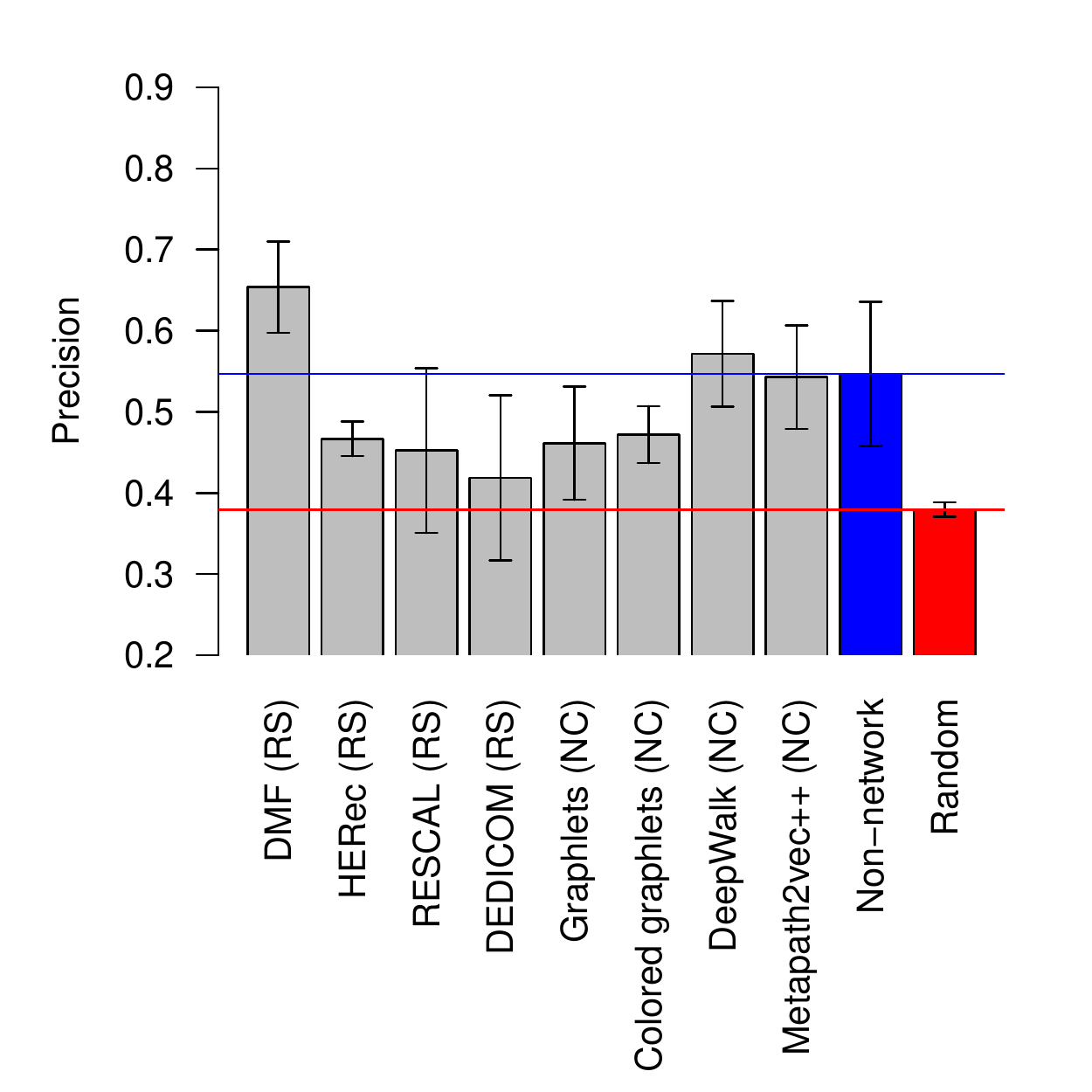}
    \captionsetup{width=\textwidth}
\caption{Anxiety}
\label{fig:5a}
\end{subfigure}
\hfill
\begin{subfigure}[b]{0.45\textwidth}
    \centering
    \includegraphics[width=\linewidth]{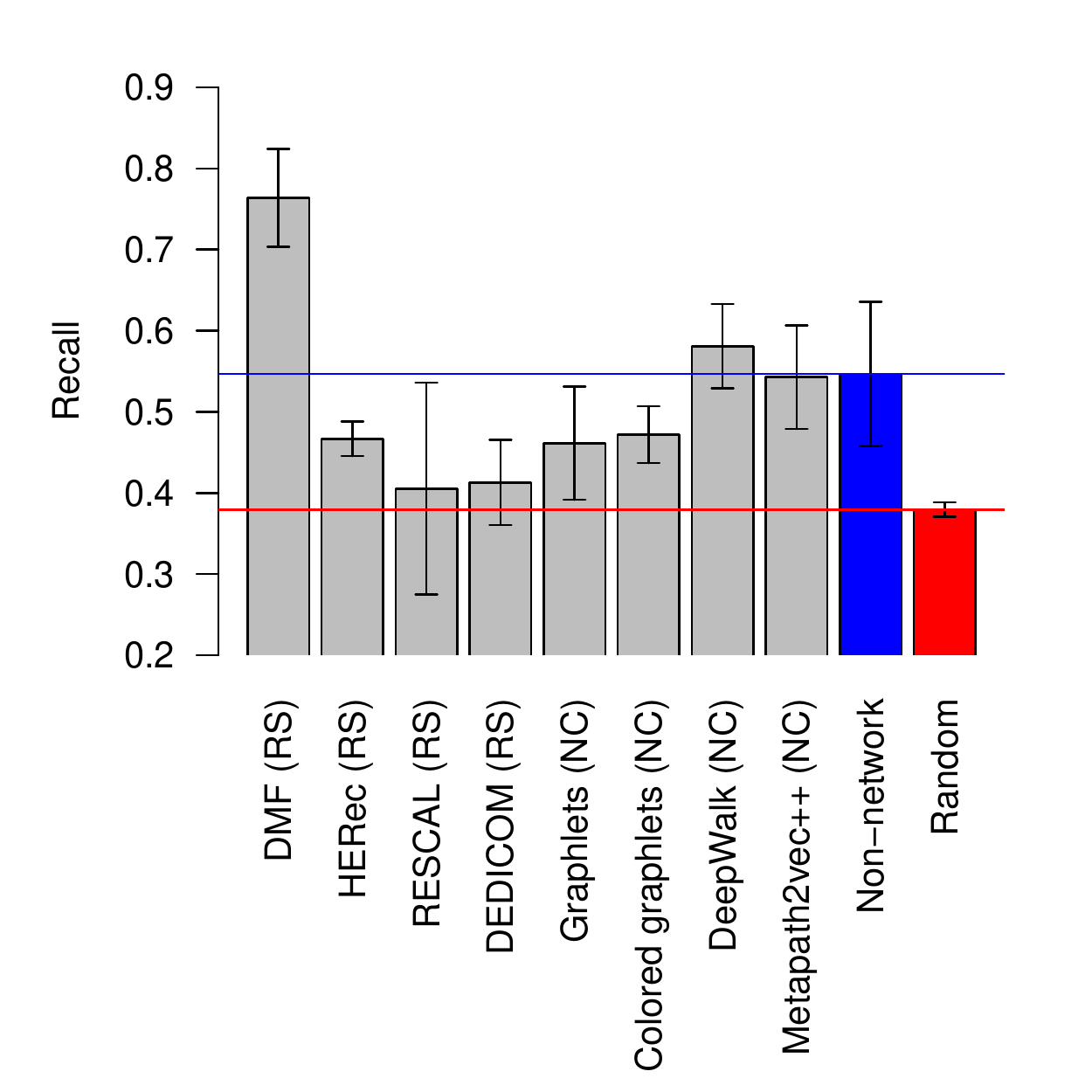}
    \captionsetup{width=\textwidth}
\caption{Anxiety}
\label{fig:5b}
\end{subfigure}

\captionsetup{width=\textwidth}
\caption{Method performance when predicting  depression and anxiety with respect to precision and recall. Each column shows the prediction performance of the corresponding method averaged over five runs of the 5-fold cross-validation; the error bar represents the corresponding standard deviation. The column corresponding to the random guess method is marked in red and its average precision or recall is shown by the red line. The column corresponding to the non-network method is marked in blue and its average precision or recall is shown by the blue line. RS denotes ``recommender system''. NC denotes ``node classification''. Analogous results for F1-score and accuracy are shown in Supplementary Figure S5 for depression and in Supplementary Figure S6 for anxiety.} \label{fig:4and5}

\end{figure}

\vspace{0.2cm}
\noindent \textit{Depression prediction.} \hspace{1mm}
Among all network methods that we evaluate, DMF, an RS method, is the most accurate and is significantly (\textit{p}-value$<$0.05) more accurate than the rest of the network methods in terms of all evaluation measures (Figures \ref{fig:4a}, \ref{fig:4b} and Supplementary Figure S5). Our results indicate that the most accurate RS method (DMF) outperforms all NC methods. However, this observation could be due to the classifier used in NC. Instead of the logistic regression classifier, using other classifiers such as random forest and support vector machine might improve the performance of NC. Recall that the reason why we have used logistic regression is that this particular classifier is typically used in the field of mental health prediction \cite{mohr2017personal,lin2016association,reece2017instagram}.

When we compare the RS and NC network methods to the random guess method, we find that all network methods, except RESCAL and DEDICOM, are significantly (\textit{p}-value$<$0.05) more accurate in terms of all evaluation measures (Figures \ref{fig:4a}, \ref{fig:4b} and Supplementary Figure S5). The most accurate method that we evaluate, DMF, achieves gains of 115\% in terms of precision, 150\% in terms of recall, 131\% in terms of F1 score, and 21\% in terms of accuracy over the random guess method. Note that when we compare an approach, say DMF, to the random guess method, we measure the gain (i.e., relative change) of DMF over the random guess method as $\frac{Performance_{\,DMF} - Performance_{\,random}}{Performance_{\,random}}$. For example, if $Performance_{\,DMF}$ is 0.612 and $Performance_{\,random}$ is 0.245, the gain is $\frac{0.612 - 0.245}{0.245}= 1.5 = 150\%$. RESCAL and DEDICOM do not accurately predict depression - RESCAL and DEDICOM show similar (i.e., not significantly different) values as the random guess method, in terms of all evaluation measures (Figures \ref{fig:4a}, \ref{fig:4b} and Supplementary Figure S5). Note that the superiority of DMF over the other two RS methods (RESCAL and DEDICOM) is not surprising: DMF was already shown to perform better than RESCAL in recommendation tasks \cite{drumond2014optimizing}. In turn, RESCAL was already shown to perform better than DEDICOM in recommendation tasks \cite{nickel2011three}. Therefore, we expected DMF to work the best of these three RS approaches. However, it is surprising that in our task of predicting mental health, RESCAL and DEDICOM produce random-like results. Also, it is at least somewhat surprising that DMF is superior than the fourth considered RS approach, HERec, given that the latter is a more recent approach than the former. Examining why  RESCAL and DEDICOM produce random-like results and why DMF is superior to HERec is non-trivial, given the heuristic-like nature of these methods in the recommendation task, without many if any theoretic guarantees. As such, this is out of the scope of the current study.

When we compare the network methods to the non-network method, we find that: 1) DMF and DeepWalk are significantly (\textit{p}-value$<$0.05) more accurate in terms of all evaluation measures (Figures \ref{fig:4a}, \ref{fig:4b} and Supplementary Figure S5); 2) graphlets, colored graphlets, and Metapath2vec++ are marginally (i.e., not significantly) more accurate in terms of all evaluation measures (Figures \ref{fig:4a}, \ref{fig:4b} and Supplementary Figure S5); 3) HERec is comparable in terms of all evaluation measures (Figures \ref{fig:4a}, \ref{fig:4b} and Supplementary Figure S5); and 4) RESCAL and DEDICOM are significantly (\textit{p}-value$<$0.05) less accurate in terms of all evaluation measures (Figures \ref{fig:4a}, \ref{fig:4b} and Supplementary Figure S5). The most accurate method that we evaluate, DMF, achieves gains (as defined above) of 41\% in terms of precision, 64\% in terms of recall, 52\% in terms of F1 score, and 11\% in terms of accuracy over the non-network method. 

In summary, our results show that for depression, the best RS method significantly outperforms all NC methods. In addition, six out of the eight network methods significantly outperform the random guess method. Moreover, both the best RS method and the best NC method significantly outperform the non-network method. This confirms the power of network methods and RS in particular in predicting depression.

\vspace{0.2cm}
\noindent \textit{Anxiety prediction.} \hspace{1mm} 
Among all network methods that we evaluate, DMF, an RS method, is the most accurate and is significantly (\textit{p}-value$<$0.05) more accurate than the rest of the network methods in terms of all evaluation measures (Figures \ref{fig:5a}, \ref{fig:5b} and Supplementary Figure S6). Our results indicate that the most accurate RS method (DMF) outperforms all NC methods.

When we compare the RS and NC network methods to the random guess method, we find that all network methods, except RESCAL and DEDICOM, are significantly (\textit{p}-value$<$0.05) more accurate than the random guess method in terms of all evaluation measures (Figures \ref{fig:5a}, \ref{fig:5b} and Supplementary Figure S6). The most accurate method that we evaluate, DMF, achieves gains of 72\% in terms of precision, 101\% in terms of recall, 86\% in terms of F1 score, and 43\% in terms of accuracy over the random guess method. RESCAL and DEDICOM do not accurately predict anxiety - RESCAL and DEDICOM show similar (i.e., not significantly different) values as the random guess method, in terms of all evaluation measures (Figures \ref{fig:5a}, \ref{fig:5b} and Supplementary Figure S6). 
 
When we compare the network methods to the non-network method, we find that: 1) DMF is significantly (\textit{p}-value$<$0.05) more accurate in terms of all evaluation measures (Figures \ref{fig:5a}, \ref{fig:5b} and Supplementary Figure S6); 2) DeepWalk is marginally more accurate and HERec, graphlets, colored graphlets, and Metapath2vec++ are marginally less accurate in terms of all evaluation measures (Figures \ref{fig:5a}, \ref{fig:5b} and Supplementary Figure S6); and 3) RESCAL and DEDICOM are significantly (\textit{p}-value$<$0.05) less accurate in terms of all evaluation measures (Figures \ref{fig:5a}, \ref{fig:5b} and Supplementary Figure S6). The most accurate method that we evaluate, DMF, achieves gains of 20\% in terms of precision, and 40\% in terms of recall, 29\% in terms of F1 score, and 16\% in terms of accuracy over the non-network method. 

In summary, the results for anxiety are similar to those for depression. This confirms the power of network methods and RS in particular in predicting anxiety in addition to depression.

\vspace{0.2cm}
\noindent \textbf{Q2: Do the most accurate RS method, the most accurate NC method, and the non-network method identify different sets of anxious/depressed individuals, i.e., are they complementary to each other?}

\begin{figure}[h!]
    \centering
    \begin{subfigure}[b]{0.4\textwidth}
    \includegraphics[width=\linewidth]{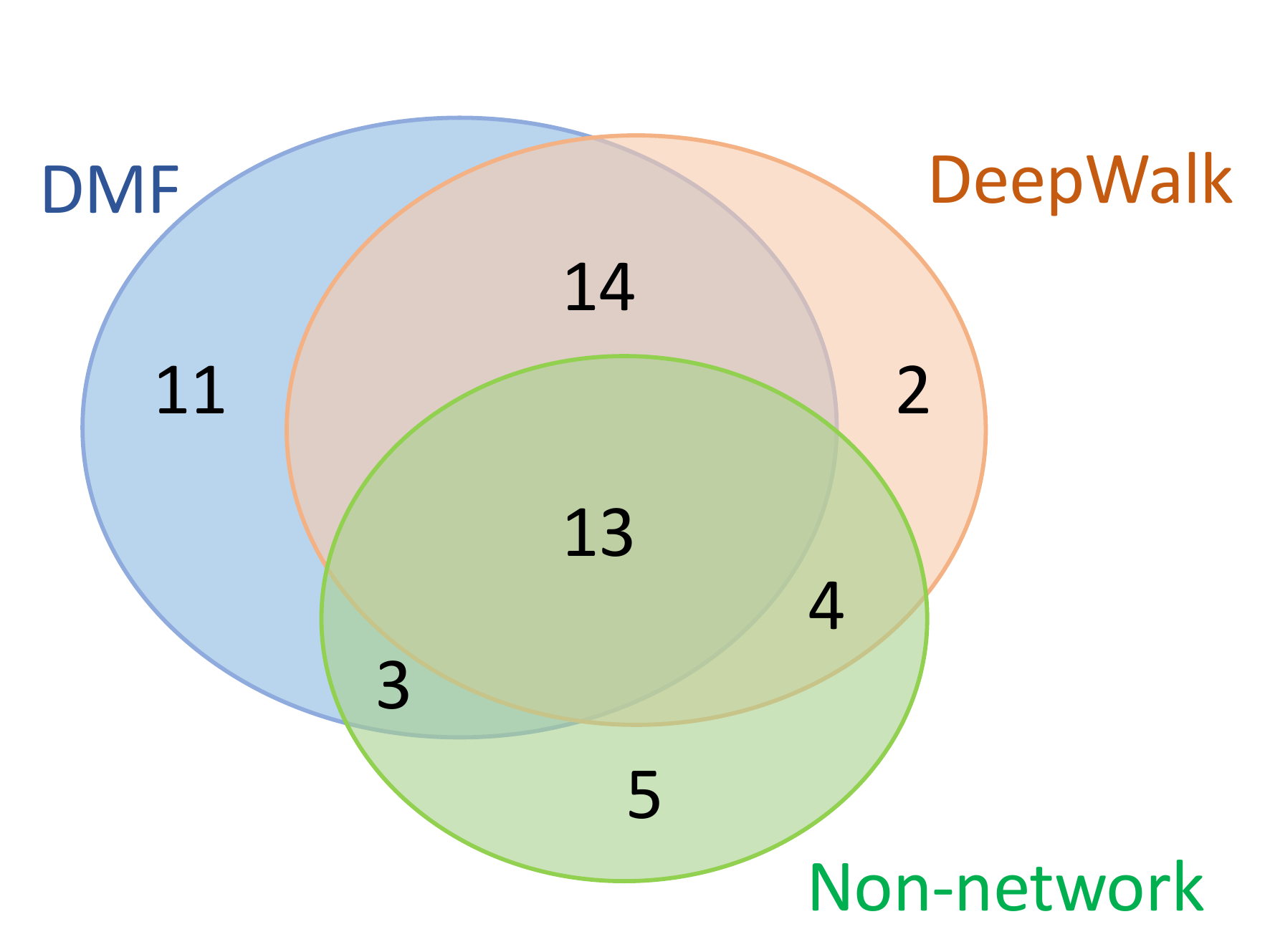}
    \captionsetup{width=\textwidth}
\caption{Depression}
\label{fig:6}
\end{subfigure}
\hfill
\begin{subfigure}[b]{0.4\textwidth}
    \centering
    \includegraphics[width=\linewidth]{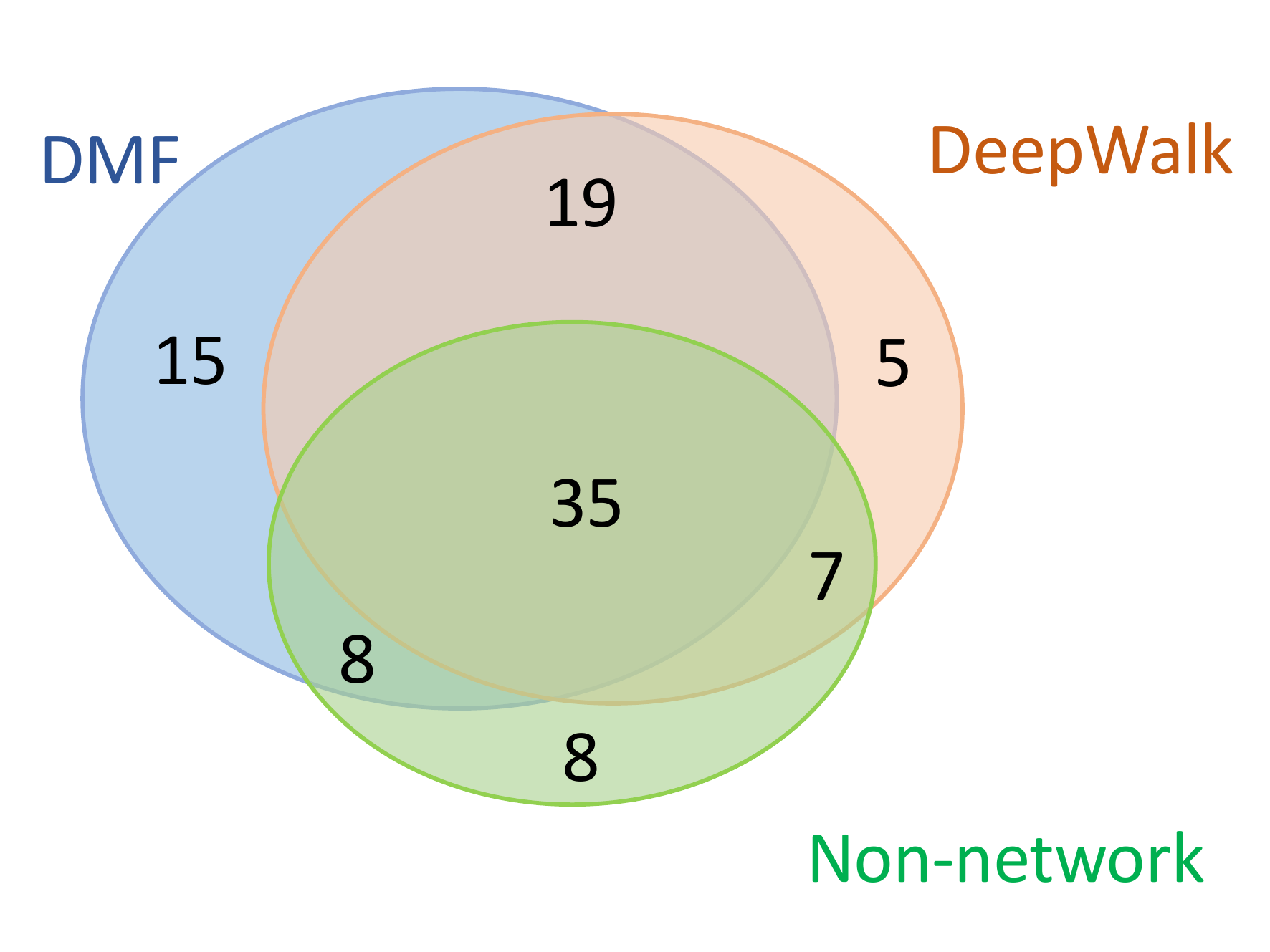}
    \captionsetup{width=\textwidth}
\caption{Anxiety}
\label{fig:7}
\end{subfigure} 
\captionsetup{width=\textwidth}
\caption{Sizes of overlaps between individuals correctly predicted as \textbf{(a) depressed} and \textbf{(b) anxious} by the best RS network method (DMF), the best NC network method (DeepWalk), and the non-network method.}

\end{figure}

To answer Q2, we examine the overlap of sets of depressed/anxious individuals correctly predicted by DMF, DeepWalk, and the non-network method. We find that among the three methods, DMF identifies the largest number of depressed/anxious individuals (Figures \ref{fig:6} and \ref{fig:7}), which is reflected by its highest prediction accuracy (Figure \ref{fig:4and5}). Using DMF, we predict as depressed 41 out of all 67 actually depressed individuals and as anxious 77 out of all 106 actually anxious individuals. The other two methods combined correctly predict additional 11 depressed individuals and 20 anxious individuals who are missed by DMF. Most of the depressed/anxious individuals are correctly predicted by at least one of the three methods; only 15 out of all 67 depressed individuals and 9 out of all 106 anxious individuals are not correctly predicted by any of the three methods.

Results regarding the overlap between predictions of DMF, DeepWalk, and the non-network method are qualitatively similar for depression and anxiety (Figures \ref{fig:6} and \ref{fig:7}). Taking depression as an example, we find that the two sets of depressed individuals correctly predicted by DMF and DeepWalk overlap significantly (\textit{p}-value$<$0.05) (Figure \ref{fig:6}). This could be because DMF and DeepWalk are both network methods and differ only in one aspect - they are different types of methods (RS and NC). Moreover, the two sets of depressed individuals correctly predicted by DeepWalk and the non-network method overlap significantly (\textit{p}-value$<$0.05) (Figure \ref{fig:6}). This could be because the two methods both use logistic regression to make predictions and differ only in one aspect - DeepWalk uses network features while the non-network method uses non-network features. However, the two sets of depressed individuals correctly predicted by DMF and the non-network method do not significantly overlap (Figure \ref{fig:6}). This could be because DMF differs from the non-network method in two ways: the former is a network method and it does not use logistic regression to make predictions, while the latter is a non-network method that uses logistic regression. In other words, it could be that the more similar the two approaches are in terms of their methodologies, the more similar their predictions. 

Since the two sets of depressed/anxious individuals correctly predicted by DMF and the non-network method do not significantly overlap, the combination of the two methods' ideas may be able to correctly predict more of the depressed/anxious individuals. Thus, we could potentially use ensemble learning algorithms to combine DMF and the non-network method to achieve more accurate predictions \cite{dietterich2002ensemble, polikar2012ensemble}.

\vspace{0.2cm}
\noindent \textbf{Q3: What is the impact of using different types of information about the individuals, represented by different edge types in the HIN, on the performance of mental health prediction?}

To answer Q3, we train a series of instances of DMF, which is the most accurate method we have evaluated thus far (when considering all types of information about individuals). Each instance is trained by using the target (i.e., {\fontfamily{cmtt}\selectfont individual -  mental health}) edge type and one or more of the other five non-target edge types. We analyze all possible combinations of the five non-target edge types. Each DMF instance uses one of the combinations. In total, we train 31 DMF instances corresponding to 31 possible combinations of the five non-target edge types. In this section, edge types are denoted as follows: 
I: {\fontfamily{cmtt}\selectfont individual -  individual}, 
P: {\fontfamily{cmtt}\selectfont individual -  personality traits}, 
S: {\fontfamily{cmtt}\selectfont individual -  social status}, 
F: {\fontfamily{cmtt}\selectfont individual -  physical health} (where ``F'' stands for mostly Fitbit-based physical health data), and 
W: {\fontfamily{cmtt}\selectfont individual -  well-being}. 
Then, the different edge type combinations are denoted by the corresponding combinations of the I, P, S, F, and W acronyms.

\begin{figure}[h!]
    \centering
    \includegraphics[width=0.8\linewidth]{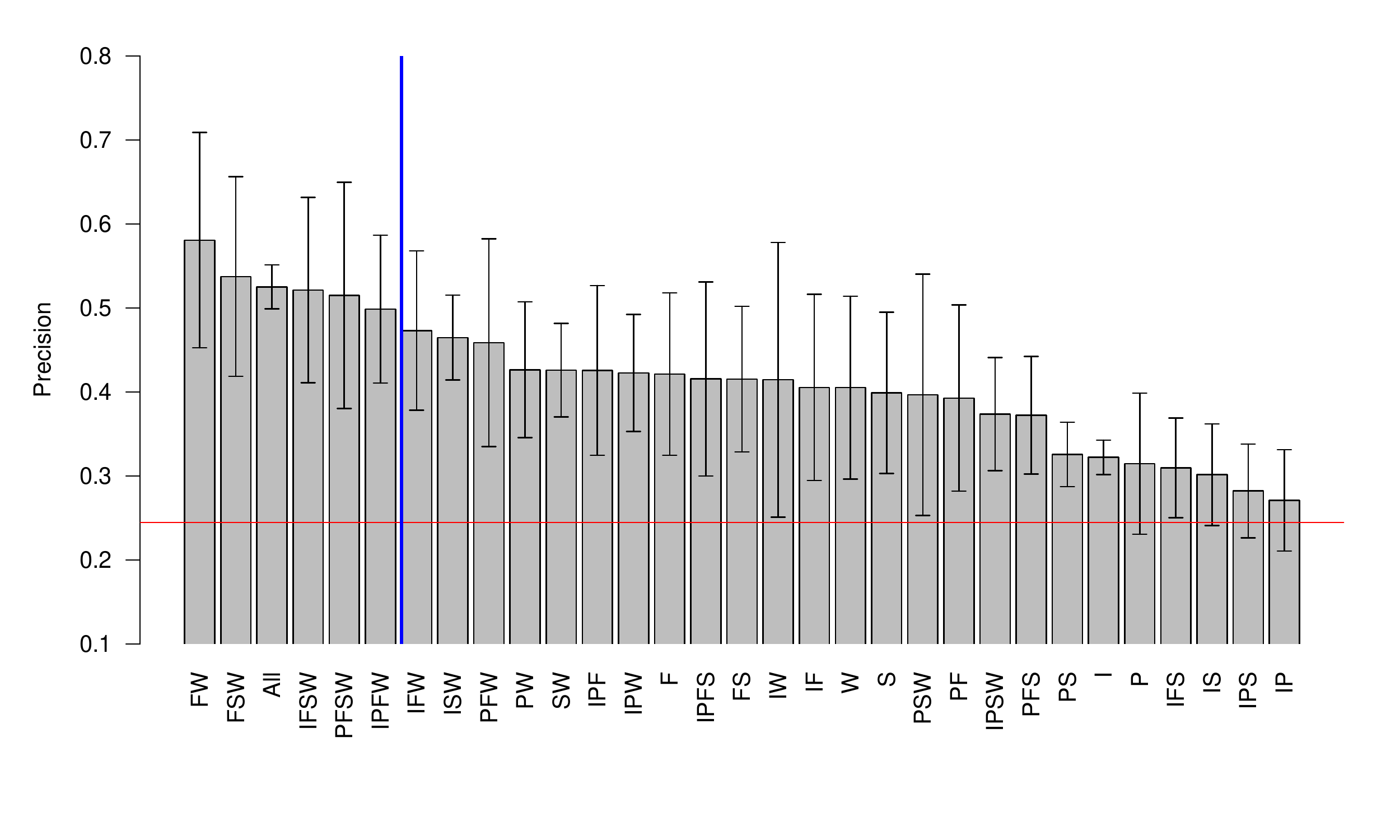}
\captionsetup{width=\textwidth}
\caption{Performance of DMF (the best network method) when using different combinations of data (i.e., edge types) to predict \textbf{depression}, with respect to precision. In this figure, edge types are denoted as follows. I: {\fontfamily{cmtt}\selectfont individual -  individual}, P: {\fontfamily{cmtt}\selectfont individual -  personality traits}, S: {\fontfamily{cmtt}\selectfont individual - social status}, F: {\fontfamily{cmtt}\selectfont individual -  physical health}, W: {\fontfamily{cmtt}\selectfont individual -  well-being}. ``All'' denotes the combination of all edge types, and it corresponds to the DMF method shown in Figure \ref{fig:4a}. Each column shows the prediction performance of the corresponding edge type combination averaged over five runs in the 5-fold cross-validation; the error bar
represents the corresponding standard deviation. Columns are sorted from left to right according to their heights from high to low, i.e., edge type combinations are sorted from left to right in decreasing order of their prediction performance. Columns to the left of the vertical blue line are the six top performing combinations that we focus on in the text. The horizontal red line shows the precision of the random guess method.}
\label{fig:8}
\end{figure}

\vspace{0.2cm}
\noindent \textit{Depression prediction.} \hspace{1mm} 
We find that among all combinations, the FW combination\textemdash representing the combination of the {\fontfamily{cmtt}\selectfont individual -  physical health} and {\fontfamily{cmtt}\selectfont individual -  well-being} (W) edge types\textemdash is the most accurate in terms of all evaluation measures (Figure \ref{fig:8} and Supplementary Figure S7). In more detail, the FW combination is significantly (\textit{p}-value$<$0.05) more accurate than the rest of combinations, including the combination of all five non-target edge types (``All''), in terms of all evaluation measures (Figure \ref{fig:8} and Supplementary Figure S7). Potential reasons why the FW combination is more accurate than the ``All'' combination are as follows. First, DMF is an MRMF method (Section \ref{sec:methods}). MRMF typically contains a large number of parameters and may be prone to overfitting, meaning that MRMF may fit well on the training data but not predict well on the testing data \cite{srivastava2014dropout}. The complexity (the number of parameters) of the ``All'' combination is higher than the complexity of the FW combination. Thus, the higher number of parameters of the former may cause its overfitting, which in turn may cause its lower prediction performance. Second, some edge types in the ``All'' combination may be less informative than the {\fontfamily{cmtt}\selectfont individual -  physical health} (F) and {\fontfamily{cmtt}\selectfont individual -  well-being} (W) edge types in the FW combination when predicting depression. Using edge types that may be suboptimally informative may lower the prediction performance compared to using only edge types that are optimally informative. In other words, it might not be surprising that using some subset of all data types might be more informative/accurate than using all data types. Specifically, in our evaluation, since each of the F and W edge types alone is more accurate than any one of S, I, and P edge types alone (Figure \ref{fig:8}), it might not be surprising that the FW edge combination is more accurate than the All combination. Importantly, it \emph{is} the case that the other edge types alone, namely S and I (although not P), are performing significantly better than at random (Figure \ref{fig:8}), meaning that they do contain some predictive power. So, it is the subject of our future work to understand how to significantly improve upon the FW combination while incorporating the S and I (and possibly even P) edge types, i.e., how to get a truly synergistic, multiplicative effect when integrating the different data types. In Section \ref{sect:conclusion}, we discuss a possible direction towards achieving this goal.

Besides FW, there exist five other combinations that all have comparable (i.e., not significantly different) prediction performance to each other and are all significantly (\textit{p}-value$<$0.05) more accurate than the rest of the combinations in terms of all evaluation measures (Figure \ref{fig:8} and Supplementary Figure S7). 
So, these five combinations can all be considered as the second best result, inferior only to the FW combination. These combinations are FSW, ``All'' (the combination including all types of edges), IFSW, PFSW, and IPFW. 
We find that all of these combinations contain the {\fontfamily{cmtt}\selectfont individual - well-being} (W) and {\fontfamily{cmtt}\selectfont individual -  physical health} (F) edge types. This agrees with the literature - well-being traits such as body image and self-esteem \cite{stice2000body, sowislo2013does}, as well as physical health traits such as physical activity \cite{harris2006physical} and sleep \cite{baglioni2011insomnia,alvaro2013systematic} are correlated with depression, which validates our HIN-based predictive framework.

\begin{figure}[h!]
    \centering
    \includegraphics[width=0.8\linewidth]{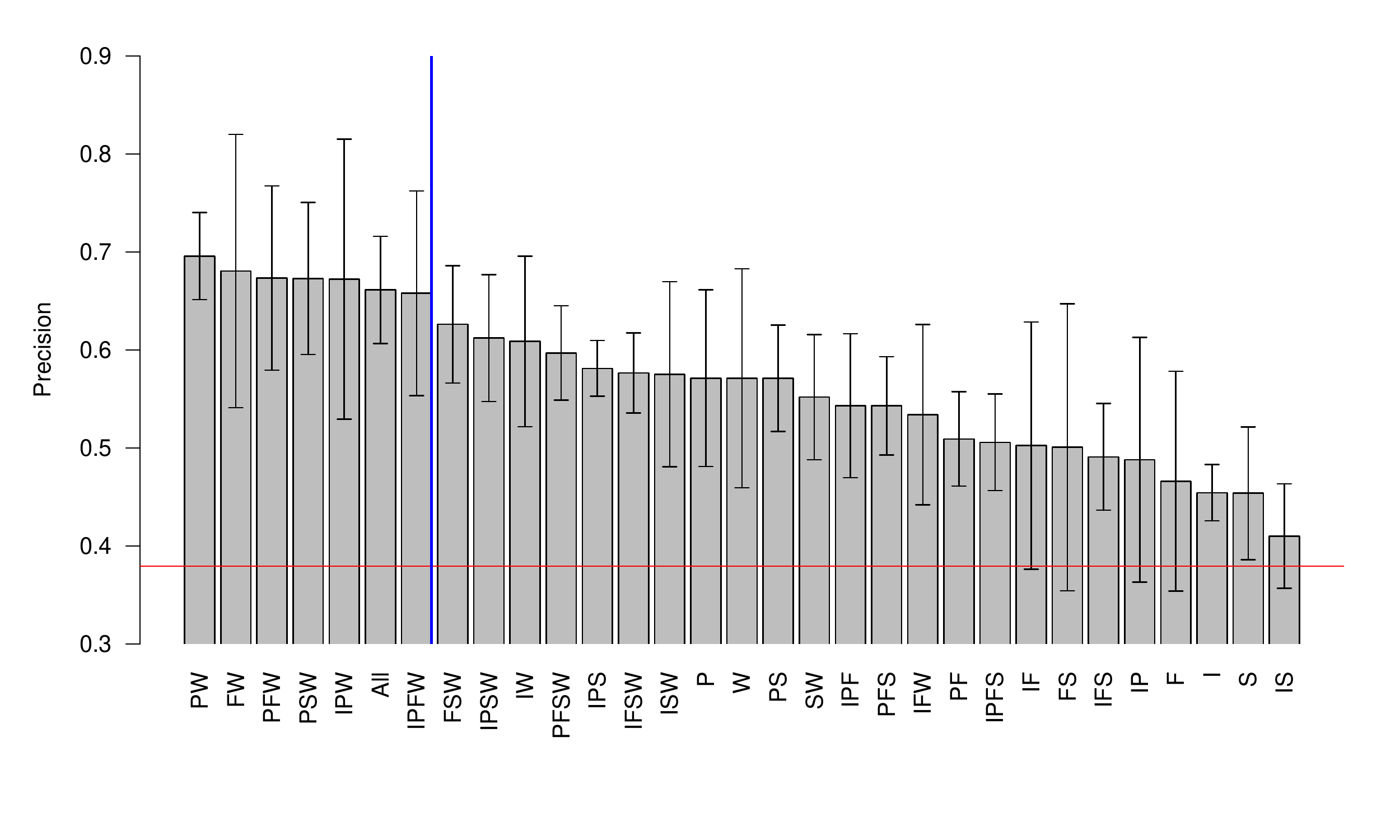}
\captionsetup{width=\textwidth}
\caption{
Performance of DMF (the best network method) when using different combinations of data (i.e., edge types) to predict \textbf{anxiety}, with respect to precision. In this figure, edge types are denoted as follows. I: {\fontfamily{cmtt}\selectfont individual - individual}, P: {\fontfamily{cmtt}\selectfont individual -  personality traits}, S: {\fontfamily{cmtt}\selectfont individual -  social status}, F: {\fontfamily{cmtt}\selectfont individual -  physical health}, W: {\fontfamily{cmtt}\selectfont individual -  well-being}. ``All'' denotes the combination of all edge types, and it corresponds to the DMF method shown in Figure \ref{fig:5a}. Each column shows the prediction performance of the corresponding edge type combination averaged over five runs in the 5-fold cross-validation; the error bar
represents the corresponding standard deviation. Columns are sorted from left to right according to their heights from high to low, i.e., edge type combinations are sorted from left to right in decreasing order of their prediction performance. Columns to the left of the vertical blue line are the seven top performing combinations that we focus on in the text. The horizontal red line shows the precision of the random guess method.}
\label{fig:9}
\end{figure}

\vspace{0.2cm}
\noindent \textit{Anxiety prediction.} \hspace{1mm} 
For anxiety, we find that among all combinations, the PW combination\textemdash representing the combination of the {\fontfamily{cmtt}\selectfont individual -  personality traits} (P) and {\fontfamily{cmtt}\selectfont individual -  well-being} (W) edge types\textemdash is the most accurate in terms of all evaluation measures (Figure \ref{fig:9} and Supplementary Figure S8). 
 Unlike for depression, the best combination for anxiety, PW, is only marginally more accurate than six other combinations including FW, PFW, PSW, IPW, ``All'' (the combination of all types of edges), and IPFW. In other words, the PW combination and the six other combination all have comparable (i.e., not significantly different) performance in terms of all evaluation measures. Plus, all seven combinations are significantly (\textit{p}-value$<$0.05) more accurate than the rest of the combinations in terms of all evaluation measures (Figure \ref{fig:9} and Supplementary Figure S8). Focusing on the seven best combinations, the {\fontfamily{cmtt}\selectfont individual -  personality traits} (P), {\fontfamily{cmtt}\selectfont individual -  well-being} (W), and {\fontfamily{cmtt}\selectfont individual -  physical health} (F) edge types are frequently included. This finding agrees with the literature that personality traits \cite{bienvenu2003personality}, well-being traits such as self-esteem \cite{sowislo2013does}, and physical health traits such as physical activity \cite{mcmahon2017physical} and sleep \cite{alvaro2013systematic} are correlated with anxiety, which further validates our HIN-based predictive framework. 

Unlike for depression, for anxiety, the All combination \emph{is} one of the best-scoring combinations. However, it is still not significantly better than the other six best-scoring combinations. So, just like for depression, for anxiety, it is also the subject of our future work to understand how to get a multiplicative effect with our HIN data integrative framework. The promise to improve is certainly there, especially because multiple individual edge types, namely P, W, and I (although not F and S) are all performing better than at random when used alone (Figure \ref{fig:9}), meaning that each of them has some predictive power. Hence, it should be possible to use these three combined, possibly also with F and S, to get the IPW combination, possibly also the All combination, to perform significantly better than the other edge combinations. Again, in Section \ref{sect:conclusion}, we propose a step in this direction.

An additional observation is that the FW combination, the most accurate combination in depression prediction, is the second-best combination in anxiety prediction. This indicates that {\fontfamily{cmtt}\selectfont physical health} (F) and {\fontfamily{cmtt}\selectfont well-being} (W) are good predictors of both depression and anxiety. On the other hand, the PW combination, the most accurate combination in anxiety prediction, is not among the best combinations in predicting depression. This indicates that {\fontfamily{cmtt}\selectfont personality traits} (P) may not be a key factor in predicting depression, while they are a good predictor of anxiety. These results suggest that depression prediction and anxiety prediction have both similarities and differences. This might be explained by a reasonably large overlap between the depressed individuals and the anxious individuals in our data. Namely, of the 67 depressed individuals and 106 anxious individuals, 51 individuals are both depressed and anxious ($p$-value$<$0.05). So, this overlap could explain the similarities between the depression prediction and anxiety prediction results. On the other hand, a number of individuals are depressed but not anxious and vice versa, which could explain the differences between the depression prediction and anxiety prediction results.


\section{Conclusion}\label{sect:conclusion}

In this paper, we integrate individuals' smartphone, wearable sensor, and survey data into an HIN and apply state-of-the-art RS and NC methods to the HIN to predict the individuals' mental health conditions. Our results indicate that among all of the network methods, DMF, an RS method, is the best, i.e., RS is better than NC as evaluated in our study. DMF outperforms the non-network method as well as the random guess method in terms of all evaluation measures. This confirms the power of network-based analyses of NetHealth data and RS in particular in predicting mental health. This study can be extended in several ways. 1) Because the NetHealth study has collected time series data, adding temporal information such as dynamic social interaction data \cite{psb2020} into our HIN could perhaps yield a truly multiplicative effect of data integration, i.e., lead to improvement compared to using the currently best-performing FW and PW data type combinations for depression and anxiety, respectively. Note that doing this is non-trivial, as traditional RS and NC methods are designed for static HINs. This is why we have not considered the data's temporal nature in the current study and why instead we plan to do so in our future work. 2) In this study, we focus on the task of mental health prediction as a proof-of-concept. But our HIN framework could be generalized to predict any of the individuals’ traits available in the NetHealth data. In other words, our framework could be generalized to predict any edge type and not just the {\fontfamily{cmtt}\selectfont individual -  mental health} edge type, as we do in the current study. To do that, we would just need to treat a desired edge type as the target edge type and the rest of the edge types (including the {\fontfamily{cmtt}\selectfont individual -  mental health} edge type) as side information. 3) Our framework could also be generalized to new and larger data sets containing more participants and more types of data when such data sets become available. To do that, we would just need to model the new data set as a new HIN that may contain a larger number of nodes and edges and more node and edge types than the HIN constructed in this study, and then apply RS and NC methods to the new HIN. 4) Because of the non-significant overlap between DMF’s and the non-network method’s predictions, ensemble learning methods could be developed to combine the two methods' ideas in order to further improve prediction performance compared to each of the methods individually. Exploring this is beyond the scope of this paper and is the subject of our future work.

\section{Acknowledgement}

This work was funded by the National Institutes of Health (NIH) 1R01HL117757, National Science Foundation (NSF) CAREER CCF-1452795, and Air Force Office of Scientific Research (AFOSR) YIP FA9550-16-1-0147 grants. We thank the entire NetHealth team for their useful discussions during weekly meetings. We especially thank the following team members: Rachael Purta for helping us access the Fitbit data, Afzal Hossain for help with the SMS data, and Louis Faust for assistance with understanding the Fitbit and survey data. Importantly, we thank all NetHealth study individuals for generously volunteering and consenting to share their data, without whom our study would not have been possible.

\bibliographystyle{abbrv}
\bibliography{main}

\end{document}